\documentclass[sn-mathphys,Numbered,iicol]{sn-jnl}

\usepackage{graphicx}%
\usepackage{multirow}%
\usepackage{amsmath,amssymb,amsfonts}%
\usepackage{amsthm}%
\usepackage{mathrsfs}%
\usepackage[title]{appendix}%
\usepackage{xcolor}%
\usepackage{textcomp}%
\usepackage{manyfoot}%
\usepackage{booktabs}%
\usepackage{algorithm}%
\usepackage{algorithmicx}%
\usepackage{algpseudocode}%
\usepackage{listings}%

\begin{document}

\title[Article Title]{Depth from Defocus Technique: A Simple Calibration-Free Approach for Dispersion Size Measurement}

\author[1]{\fnm{Saini Jatin} \sur{Rao}}\email{jatinrao@iisc.ac.in}

\author[1]{\fnm{Shubham} \sur{Sharma}}\email{shubhams12@iisc.ac.in}

\author*[1,2]{\fnm{Saptarshi} \sur{Basu}}\email{sbasu@iisc.ac.in}

\author*[3]{\fnm{Cameron} \sur{Tropea}}\email{ctropea@sla.tu-darmstadt.de}

\affil[1]{\orgdiv{Department of Mechanical Engineering}, \orgname{Indian Institute of Science}, \orgaddress{ \city{Bengaluru}, \postcode{560012},  \country{India}}}

\affil[2]{\orgdiv{Interdisciplinary Centre for Energy Research}, \orgname{Indian Institute of Science}, \orgaddress{ \city{Bengaluru}, \postcode{560012},  \country{India}}}

\affil[3]{\orgdiv{Institute of Fluid Mechanics and Aerodynamics}, \orgname{Technical University of Darmstadt}, \orgaddress{ \city{Darmstadt}, \postcode{64287},  \country{Germany}}}

\abstract{Particle size measurement is crucial in various applications, be it sizing droplets in inkjet printing or respiratory events, tracking particulate ejection in hypersonic impacts, or detecting floating target markers in free surface flows. Such systems are characterised by extracting quantitative information like size, position, velocity and number density of the dispersed particles, which is typically non-trivial. The existing methods like phase Doppler or digital holography offer precise estimates at the expense of complicated systems, demanding significant expertise. We present a novel volumetric measurement approach for estimating the size and position of dispersed spherical particles that utilises a unique ‘Depth from Defocus’ (DFD) technique with a single camera. The calibration free sizing enables in-situ examination of hard to measure systems, including naturally occurring phenomena like pathogenic aerosols, pollen dispersion or raindrops. The efficacy of the technique is demonstrated for diverse sparse dispersions, including dots, glass beads, spray droplets, and pollen grains. The simple optical configuration and semi-autonomous calibration procedure make the method readily deployable and accessible, with a scope of applicability across vast research horizons.}




\maketitle

\section{Introduction}
\label{sec:Introduction}

Dispersions are heterogeneous mixtures of particles dispersed within a continuous phase, whereby the term 'particle' can refer to particles of any phase, e.g. drops/aerosols, bubbles or solid particles. 
 These particulate systems are omnipresent and they bear significance in numerous natural and practical applications. For instance, in industrial settings, the size, location and velocity of atomized fuel droplets are crucial for evaporation, rapid ignition and achieving higher efficiency of combustion based engines. Parallel examples apply to the pharmaceutical, food, agriculture, energy, and automobile industries. Understanding the transport mechanism of toxic dispersions, such as contagious aerosol droplets, dust, or microplastics, is crucial for health care and environmental sciences since it is constrained by their size ranges. From a biological perspective, entities such as pollen, blood cells, vesicles, or microorganisms possess characteristics that are dependent on their size. The list is endless, but in summary, the need to characterise the size, position and velocity of dispersed particles in a mixture is ubiquitous. Knowing such information then also allows for concentration and flux to be measured. 

Among the numerous alternatives to perform such measurements, optical methods are of particular interest, as they are non-intrusive. Optical techniques are usually characterised as being pointwise, planar or volumetric and are based on various principles, such as interferometry (e.g. phase Doppler, holography, laser diffraction, ILIDS/IPI, etc.), time shift, or direct imaging  \citep{tropea_optical_2011}. However, pointwise or planar methods are tedious to deploy when volumetric information is required for two reasons. For one, the measurement point or plane must be traversed throughout the flow field,  necessitating tedious measurement repetition and demanding steady flow conditions during the entire measurement procedure. Furthermore, the measurement volume is seldom known exactly, making a quantitative computation of global volumetric distributions difficult. Holography offers a volumetric measurement and furthermore, in-line holography is optically quite simple to realize. Nevertheless, holography does involve considerable computational effort, making the processing time longer. 

\begin{figure*}[ht]
\centering
\includegraphics[width=0.7\linewidth]{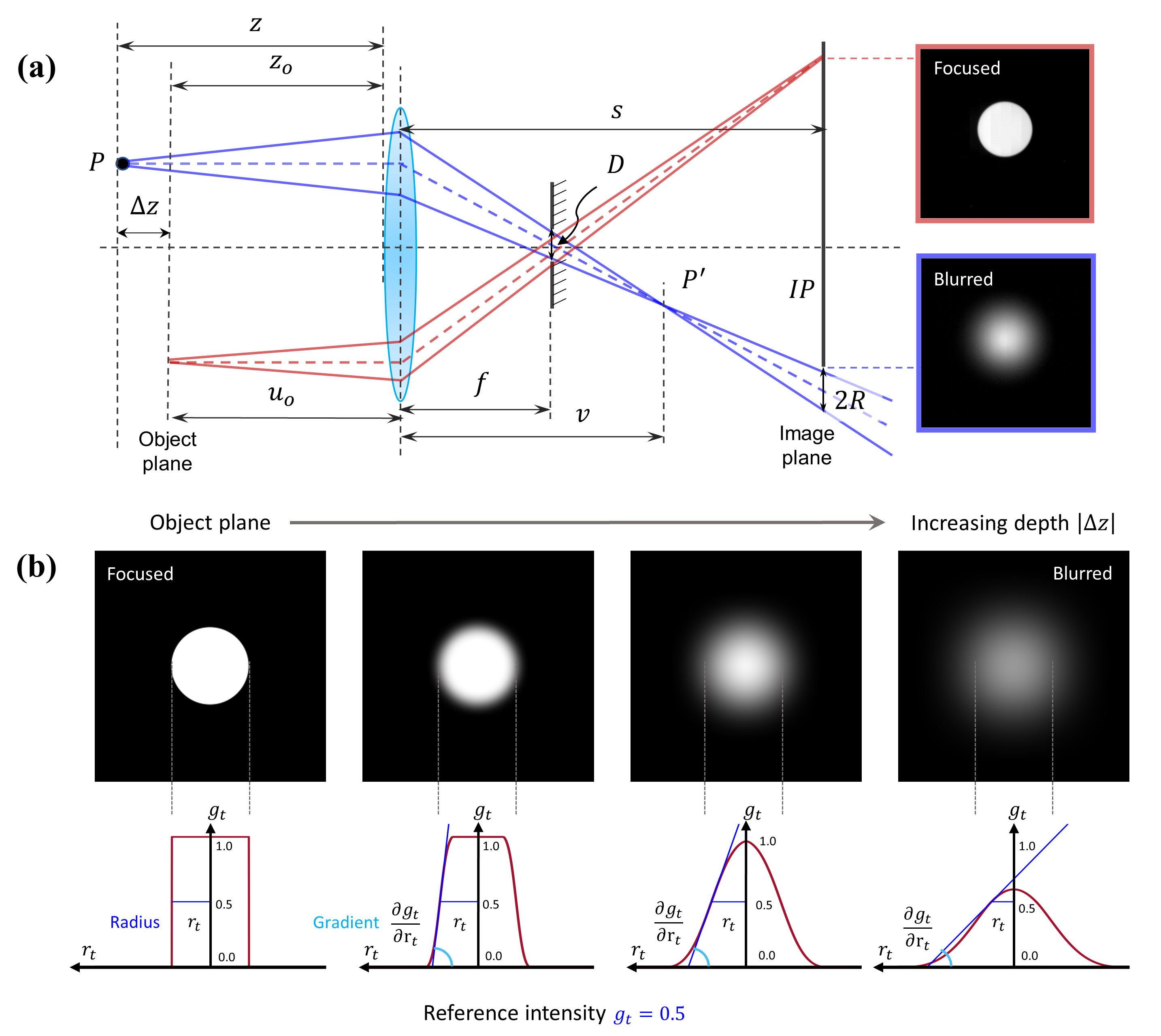}
\caption{(a) Illustration of image projection using ray optics, where a particle located at a distance $u_o$ from the lens in the object plane is in focus at a distance $s$ in the image plane (IP). Objects in front or behind the object plane by a distance $|\Delta z|$ appear blurred on the image plane. (b) Graphical illustration of particle size estimation using single camera image by extracting two quantities – radius ($r_t$) and intensity gradient ($\partial g_t/\partial r_t$) at a reference intensity value ($g_t$=0.5); both of which decrease with increasing depth from object plane $|\Delta z|$.}
\label{fig:blurring}
\end{figure*}

Direct imaging techniques provide a potential solution as they allow for high spatiotemporal resolution combined with simple experimental configurations. Shadow imaging is one such favourable configuration suitable to distinguish the particulate content from the continuous phase and furthermore, it is easy to set up and adjust \citep{erinin_comparison_2023}. However, delineating the observation volume is difficult with such approaches. As the particle moves out of focus away from the object plane (Fig.~\ref{fig:blurring}a), projected geometric features become blurred and the apparent size seems to increase. Hence, most of the early implementations of direct imaging involved only the measurement of particles in focus and rejection of the blurred projections based on grey level intensity \citep{fantini_drop_1990}, gradient \citep{hay_backlighted_1998,lecuona_volumetric_2000} or contrast based criteria \citep{kim_drop_1994}. In many applications, near focus instances occur less often, resulting in a small sample size and consequently increasing the statistical uncertainty of the measurement. Moreover, smaller particles tend to blur more rapidly with increasing distance from the object plane, reaching beyond the detection limit faster than the larger particles. This leads to an intrinsic bias in evaluating the size distribution using arithmetic averaging by overweighting the occurrence of larger particles.

These drawbacks can be mitigated by considering volumetric methods where the blurring of the out-of-focus particles is utilized to determine not only size, but also position through the degree of blurring. Such systems are known as the Depth from Defocus (DFD) approach, first introduced in the context of general imaging systems \citep{pentland_new_1987, krotkov_focusing_1988}. Several extensions were then proposed, which can be broadly classified into single or two image approaches. The single image approach is realised through special apertures \citep{willert_three-dimensional_1992,levin_image_2007,cao_defocus-based_2019}, lenses \citep{cierpka_simple_2010} or active illumination \citep{ghita_video-rate_2001}. Another approach is to employ image processing algorithms based on the concept of deconvolution \citep{ens_investigation_1993,subbarao_depth_1994}, normalised contrast \citep{blaisot_droplet_2005, fdida_drop_2010}, circle of confusion \citep{legrand_single_2016} or even machine learning \citep{saxena_3-d_2008, wang_characterization_2022, wang_three-dimensional_2022}. Some of these methods offer both size and depth estimation, albeit with an ambiguity in the depth direction, as blurring is symmetric across the object plane. Furthermore, these methods usually require a lengthy calibration procedure. 

The two image DFD approach involves acquiring the images at different degrees of blur (out of focus). This can be realised from a single camera by capturing sequential images after changing the parameters of the optical system \citep{subbarao_focused_1995} or by using coloured illumination with suitable filters \citep{murata_particle_1999}. Alternatively, two cameras and a beam splitter can be deployed to obtain simultaneous images, each with a different degree of focus. Recent developments of the two camera DFD \citep{zhou_spray_2020, zhou_sensitivity_2021,sharma_depth_2023} enable reliable measurement of size and depth using images from two cameras, whose object planes have a prescribed spacing. These images are processed using functions determined from the calibration procedure, requiring a series of target dot images of known size moved along the optical axis at known depths. Unlike other methods, this DFD approach enables the precise estimation of the measurement volume (or more precisely, the detection volume), which varies with particle size. The theoretical formulation of this two camera DFD \citep{sharma_depth_2023} lays the foundation for the present newly proposed technique using only one camera for the measurement of spherical particles.

The underlying principle of the proposed technique is illustrated in Fig.~\ref{fig:blurring}b. When the dispersed particle of interest is located on the object plane of the lens, a focused image with distinct features is obtained. However, as the particle is displaced along the depth axis away from this plane, blurring occurs, resulting in smoother features and lower intensity gradients. Another parameter of interest is the thresholded radius of the particle, which decreases as the depth increases. In the earlier DFD approaches, two experimental calibration functions were employed, utilizing the radius information obtained from two cameras for analysis. Since the actual size and position of a particle also influence the gradient magnitude of its projection, it is utilized in the single camera approach proposed here. This approach aims to determine the size and depth of a particle using thresholded radius and gradient magnitude extracted from a single image at a reference intensity (chosen here as 0.5). This is achieved using the analytical calibration functions.

\begin{figure*}[ht]
\centering
\includegraphics[width=0.9\linewidth]{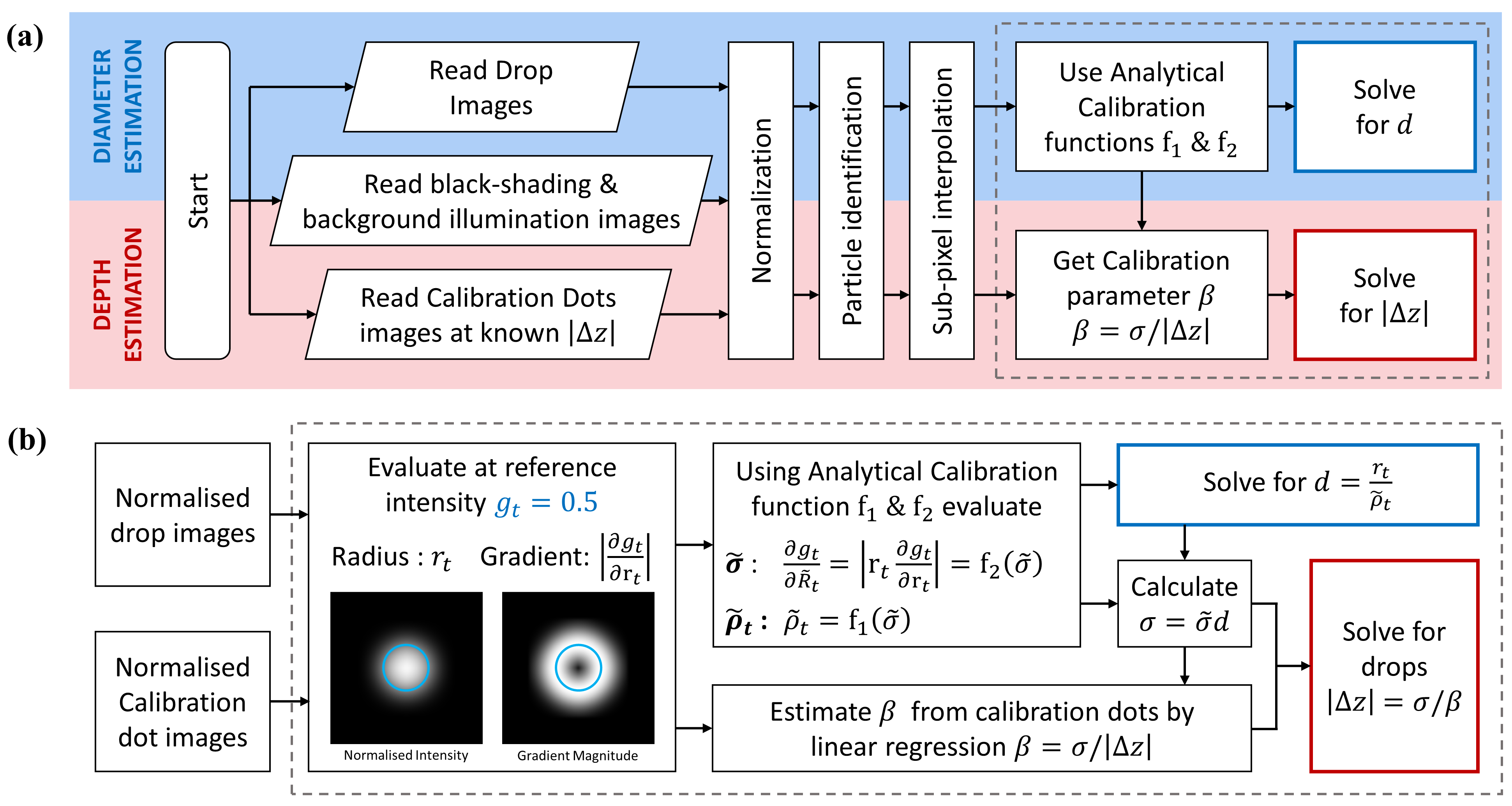}
\caption{(a) Image processing flow chart depicting the calibration free diameter estimation and depth estimation based on calibration from target dot images. (b) Expanded flow chart for the dashed boxed part in (a).}
\label{fig:flowchart}
\end{figure*}

\section{Theoretical Analysis}
\label{sec:Theoretical Analysis}
This novel implementation of a single image DFD relies on a theoretical description of the image blurring as a function of the position of the dispersed particle with respect to the object plane of the system. How this description is implemented into the calibration and into a practical measurement procedure is summarized graphically in  Fig.~\ref{fig:flowchart}. However, the imaging processing is always performed on a normalized grayscale, where the intensity values are scaled between 0 and 1. This normalization step and the image processing algorithm are described in more detail in the \textbf{Materials and Methods} (Section~\ref{sec:Materials and Methods}).

\subsection{Blurred Image Formation}

The image projection onto a camera sensor can be described using simple ray optics, as illustrated in Fig.~\ref{fig:blurring}a. When a particle is on the object plane at a distance $u_o$ from the lens, a focused image is formed at the imaging plane, located at a distance $s$ from the lens. However, when the particle is displaced to a distance $|\Delta z|$ from the object plane, the focused image shifts to a different plane, causing a blurred image projection on the sensor. This blurred image can be described by a convolution of the focused image ($i_f$) of a particle (size $d_0$) with a blurring kernel ($h$) \citep{blaisot_droplet_2005, zhou_spray_2020}. The intensity $g_t$ at any location $r_t$ is then evaluated as (Fig.~\ref{fig:convolution}a):

\begin{figure}[ht]
\centering
\includegraphics[width=0.8\linewidth]{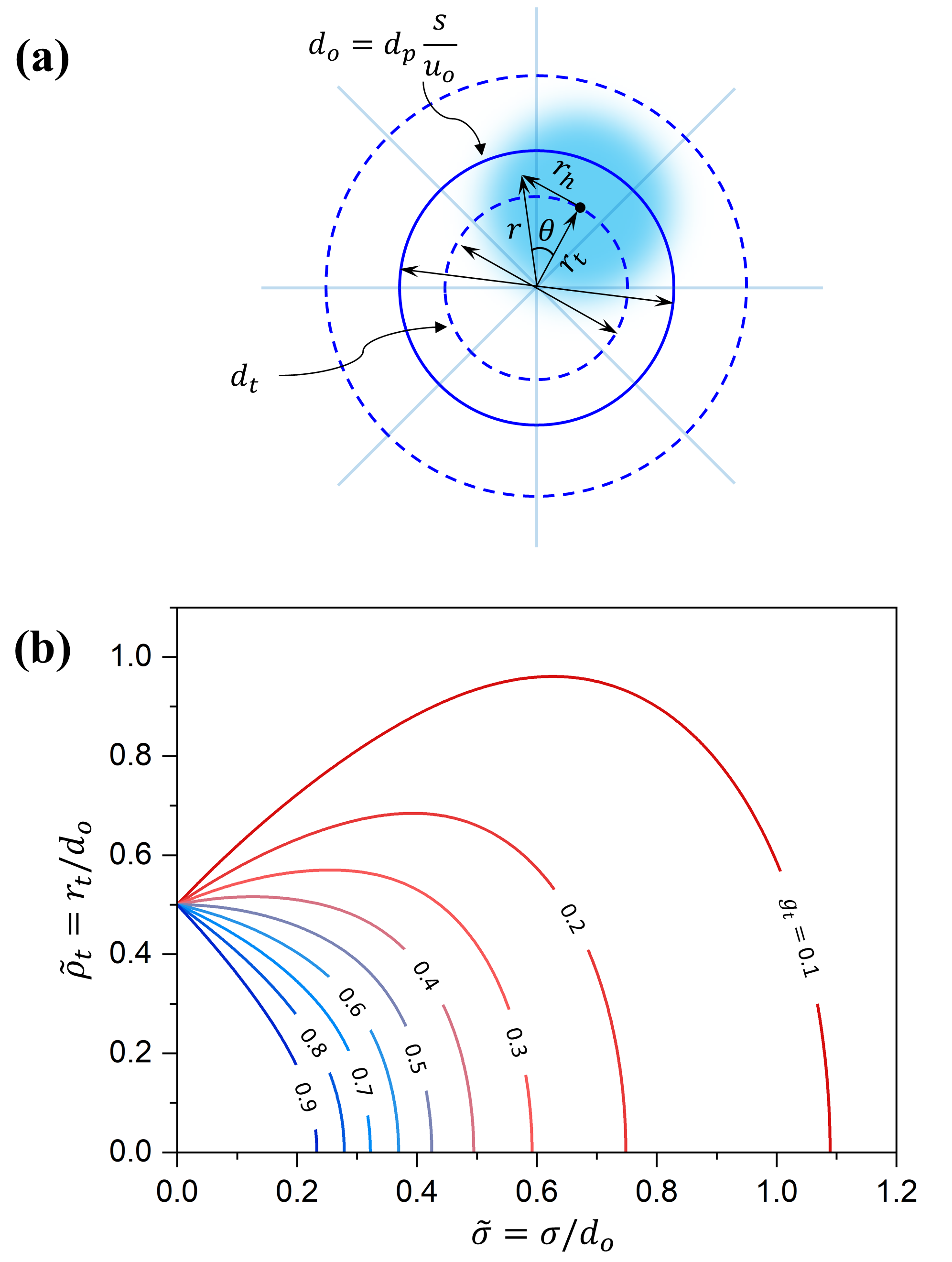}
\caption{(a) The blurred image is estimated by convolving the focused image of a particle of size $d_o$ with a Gaussian blur kernel (shown as a shaded circle). The intensity ($g_t$) at each location ($r_t$) is evaluated by convoluting the focused image with the point spread function. (b) Theoretical variation of dimensionless parameters $\tilde{\rho}_t = r_t/d_o$ with $\tilde{\sigma} = \sigma/d_o$ for different intensity threshold values ($g_t$ = 0.1 to 0.9).}
\label{fig:convolution}
\end{figure}

\begin{equation} \label{eq:convolution}
g_t(r_t) = i_f(r) \ast h (r_h)   
\end{equation}
\\
Here $i_f(r)$ is a normalized intensity image of a particle of radius $r_o = d_o/2$ on the image plane in polar coordinates 
\begin{equation} \label{eq:f_r}
\begin{split}
i_f(r) & = 0,    \quad\quad   \textrm{if}    \quad\quad   r >r _o \\
& = 1,         \quad\quad     \textrm{if}    \quad\quad  0 < r < r _o
\end{split}
\end{equation}
\\
The particle dimension on the image plane $d_o$ is related to the actual size $d_p$ through $d_o = d_p \times M$, where $M$ is the magnification of the optical system. The blur kernel $h(r_h)$ can be represented using a Gaussian profile with $\sigma$ as the standard deviation:

\begin{equation} \label{eq:h_r}
h(r_h) = \frac{1}{2 \pi \sigma^2} e^{-{\frac{{r_h}^2}{2\sigma^2}}} = \frac{1}{2 \pi \sigma^2} e^{-{\frac{(r^2+{r_t}^2-2rr_tcos\theta)}{2\sigma^2}}}
\end{equation}
\\
where $\Vec{r_h} = \Vec{r}-\Vec{r_t}$. Therefore, the two-dimensional convolution Eq.~(\ref{eq:convolution}) can be written as:

\begin{equation} \label{eq:g_rt}
g_t(r_t) =  \int_{0}^{2\pi} { \int_{0}^{d_o/2} \frac{1}{2 \pi \sigma^2} e^{-{\frac{(r^2+{r_t}^2-2rr_tcos\theta)}{2\sigma^2}}} \,rdr }\,d\theta \ 
\end{equation}
\\
The standard deviation $\sigma$ represents the degree of blur or size of the blur kernel, which can be expressed as \citep{zhou_spray_2020}

\begin{equation} \label{eq:sig-delz}
\sigma=\frac{ADM}{2f} \lvert \Delta z \rvert = \beta\lvert \Delta z \rvert
\end{equation}
\\
where $A$ is an experimental constant for the imaging system, $D$ is the aperture diameter, $f$ is the focal length and $\Delta z$ is the distance of the particle from the object plane (see Fig.~\ref{fig:blurring}a). As $A, D, M$ and $f$ are invariant for a given DFD measurement system, these terms are replaced with a single constant  $\beta$. Ultimately the resolution of imaging systems is limited by diffraction, and the smallest possible point spread function (PSF) is associated with the formation of the Airy disk. This limits the contour sharpness when in focus i.e. $\sigma \neq 0$ at $\Delta z = 0$. However, for the present system parameters, this  diffraction limitation is negligible, and other factors are more prominent, as discussed in detail in the \textbf{Appendix~\ref{Appendix-A1}}.

\begin{figure}[ht]
\centering
\includegraphics[width=0.8\linewidth]{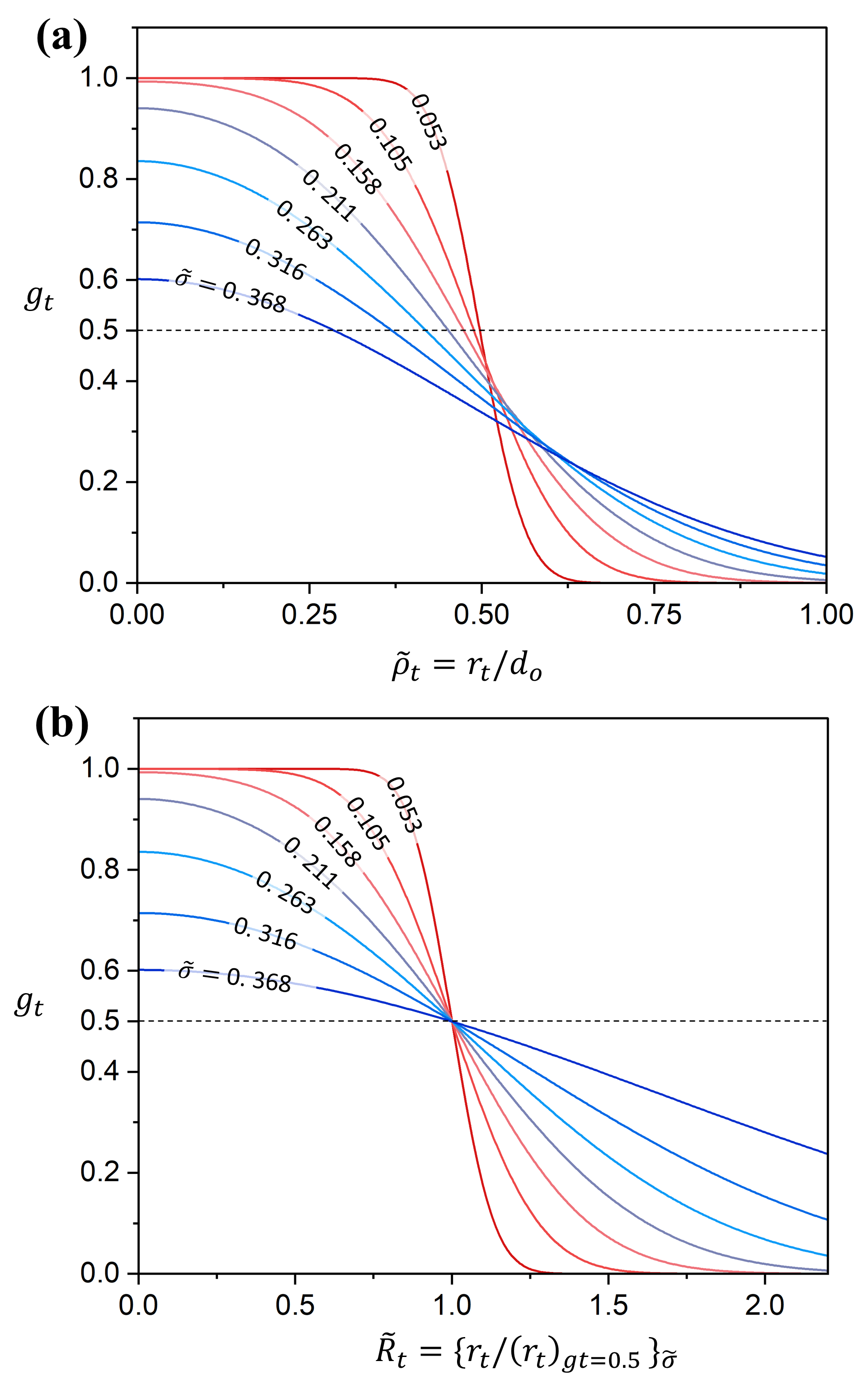}
\caption{(a) Analztical variation of intensity $g_t$  with dimensionless radius $\tilde{\rho}_t$ for various dimensionless blurring standard deviations $\tilde{\sigma}$ which is representative of depth} (b) Theoretical intensity variation with modified dimensionless radius $\tilde{R}_t$ for various dimensionless blurring standard deviations $\tilde{\sigma}$.
\label{fig:convolution_sol}
\end{figure}

The solution for the convolution integral equation Eq.~(\ref{eq:g_rt}) is obtained by non-dimensionalisation of the variables with the particle diameter as
\begin{equation} \label{eq:basic-nondim}
\widetilde{\rho}=\frac{r}{d_o},\ \ \widetilde{\rho_t}\ =\frac{r_t}{d_o},\ \ \widetilde{\sigma}=\frac{\sigma}{d_o}
\end{equation}
\\
Here we use $\widetilde{\sigma}$ as a parameter to represent the dimensionless depth from the object plane; refer Eq.~(\ref{eq:sig-delz}). Using appropriate substitutions in Eq.~(\ref{eq:g_rt}) the reduced dimensionless form is obtained as
\begin{equation} \label{eq:nondim-g_rt}
g_t\left(\widetilde{\rho_t}\right)=\frac{1}{{\widetilde{\sigma}}^2}\int_{0}^{1/2}{e^{-\frac{{\widetilde{\rho}}^2+{{\widetilde{\rho}}_t}^2}{2{\widetilde{\sigma}}^2}}I_o\left(\frac{\widetilde{\rho}\widetilde{\rho_t}}{{\widetilde{\sigma}}^2}\right)\widetilde{\rho}d\widetilde{\rho}}
\end{equation}
\\
where $I_o$ is the zeroth order modified Bessel function of the first kind. The dimensionless equation, Eq.~(\ref{eq:nondim-g_rt}) is the foundation for the analytical calibration functions, the solutions of which are numerically determined and are depicted in Figs.~\ref{fig:convolution}b and \ref{fig:convolution_sol}a. One must note that when $\widetilde{\sigma} \rightarrow 0$, the term inside the Bessel function and hence, the function itself blows up in Eq.~(\ref{eq:nondim-g_rt}). To obtain a solution in this region, the asymptotic estimate of the function as $({\widetilde{\rho}\widetilde{\rho_t}}/{{\widetilde{\sigma}}^2}) \rightarrow \infty$ is used \citep{bowman_introduction_2012}. Fig.~\ref{fig:convolution}b represents the variation of threshold radius with particle depth from the object plane for a specific threshold intensity. This solution also provides the foundation for the calibration curves used in the earlier two camera DFD approach \citep{sharma_depth_2023}. Fig.~\ref{fig:convolution_sol}a represents the intensity distribution of blurred images in the radial direction at a specific depth.

\begin{figure}[ht]
\centering
\includegraphics[width=0.8\linewidth]{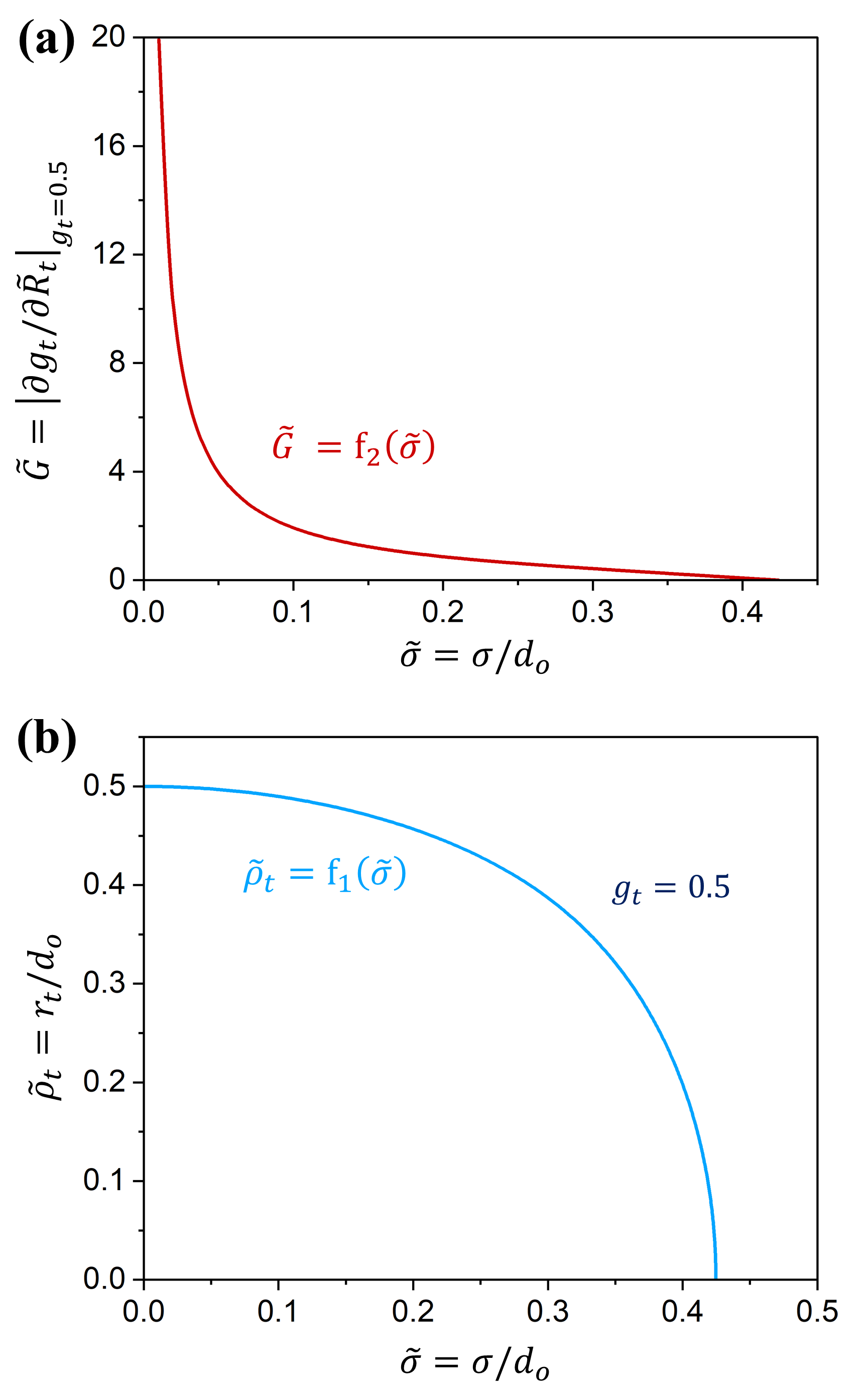}
\caption{Analytical calibration curves $f_1$  and $f_2$ for reference intensity $g_t$=0.5. (a) Variation of dimensionless gradient $\tilde{G}$ with depth, i.e., $\tilde{G}$ = $f_2(\tilde{\sigma})$ (b) Variation of dimensionless radius $\tilde{\rho}_t$ with depth, i.e., $\tilde{\rho}_t$ = $f_1(\tilde{\sigma})$.}
\label{fig:calibration}
\end{figure}

\subsection{Analytical Calibration Functions}
\begin{figure}[ht]
\centering
\includegraphics[width=0.8\linewidth]{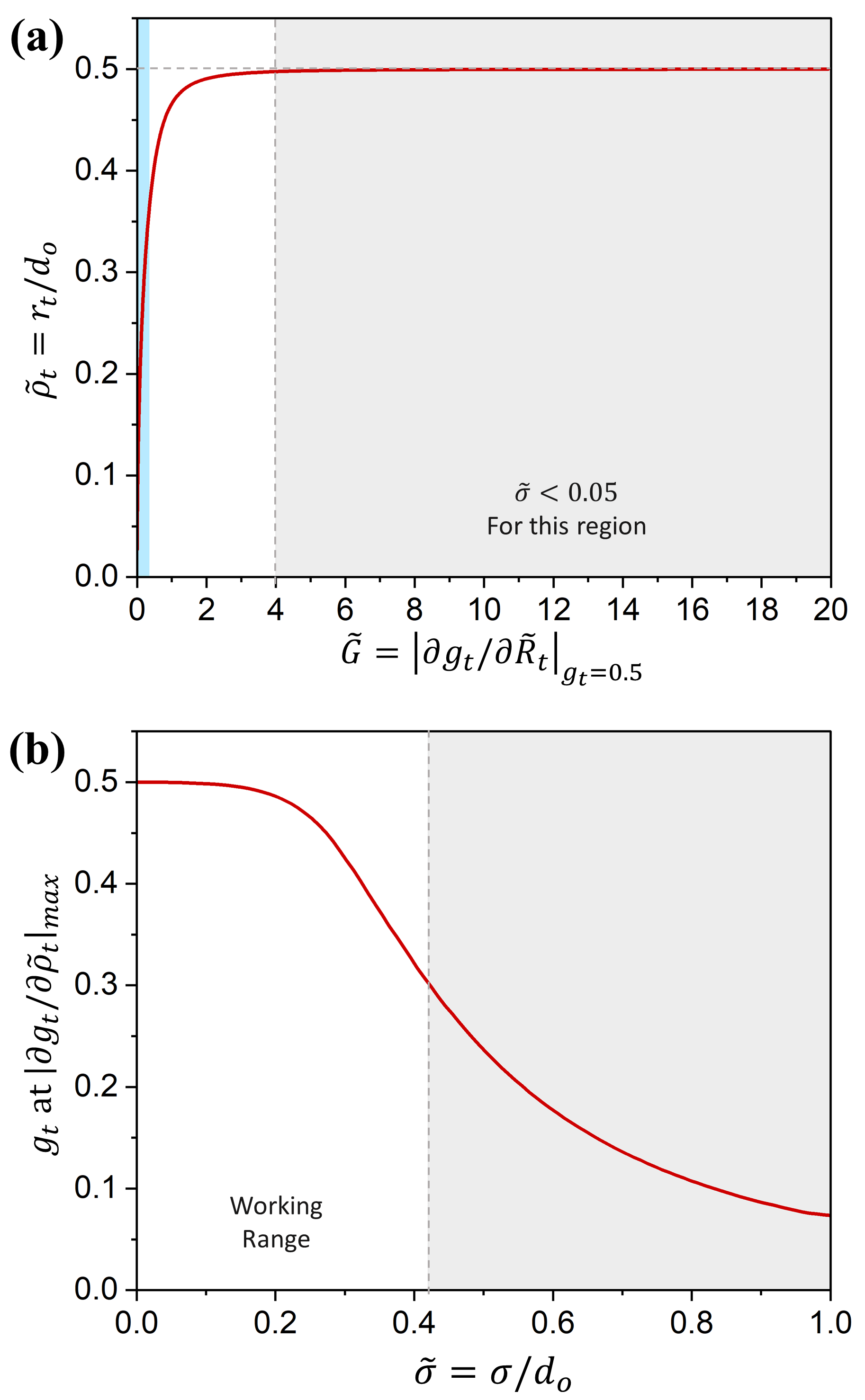}
\caption{(a) Combination of analytical calibration curves ${\tilde{\rho}}_t=f_1\left(f_2^{-1}\left(\tilde{G}\right)\right)$ for the reference intensity $g_t$=0.5. A steep variation of $\tilde{\rho}_t$  with the gradient is observed in the blue shaded region where $\tilde{\sigma}>0.35$. At the same time, there is a minimal variation in the grey shaded region, i.e., $\tilde{\rho}_t \approx 0.5$, where $\tilde{\sigma}<0.05$ (b) Variation of the intensity value at the location of maximum gradient magnitude with dimensionless depth $\tilde{\sigma}$. This corresponds to $g_t \approx 0.5$ for most of the suitable working range ($\tilde{\sigma} \leqslant 0.2$) and therefore is chosen as reference. Beyond the working range, i.e., the grey shaded region $r_t \rightarrow 0$, as is evident from calibration function $f_1$.}
\label{fig:limits}
\end{figure}

From the single camera image, two quantities can be extracted – radius $\left(r_t\right)$ and intensity gradient $\left(\partial g_t/\partial r_t\right)$ at a reference intensity value $g_t=0.5$. These parameters decrease with increasing depth of the particle from the object plane $\left|\Delta z\right|$ (Fig.~\ref{fig:blurring}b), indicating the possibility of a gradient based calibration function to estimate the degree of blur; hence, indirectly the depth. This is confirmed in Fig.~\ref{fig:convolution_sol}a by observing the intensity profiles for blurred particles at different depths, exhibiting different gradients at a reference intensity. Using an experimental image, we can only evaluate the radial intensity profiles, i.e., $r_t$-$g_t$ variation rather than the dimensionless version shown in Fig.~\ref{fig:convolution_sol}a, since $d_o$ is unknown. Hence, we propose a novel measurable dimensionless radius: 
\begin{equation} \label{eq:nondim-rad}
{\widetilde{R}}_t=\left(\frac{{\widetilde{\rho}}_t}{\left({\widetilde{\rho}}_t\right)_{g_t=0.5}\ }\right)_{\widetilde{\sigma}}=\left(\frac{r_t}{\left(r_t\right)_{g_t=0.5\ }}\right)_{\widetilde{\sigma}}
\end{equation}
\\
where $\left(r_t\right)_{g_t=0.5}$ is the radius at the reference intensity. The corresponding modified solution is depicted in Fig.~\ref{fig:convolution_sol}b, and the proposed functional form of the calibration function based on the modified gradient at reference intensity $g_t=0.5$ is

\begin{equation} \label{eq:calfun2}
\widetilde{G}\ =\left|\frac{\partial g_t}{\partial{\widetilde{R}}_t}\right|_{g_t=0.5}=\left|r_t\frac{\partial g_t}{\partial r_t}\right|_{g_t=0.5}=f_2\left(\widetilde{\sigma}\right)
\end{equation}
\\
From this measurable dimensionless version of intensity gradient $|\partial g_t/\partial {\widetilde{R}}_t| = |r_t \partial g_t/\partial r_t|$ at the reference intensity (subscript $g_t=0.5$ omitted for brevity from now on), we can estimate the dimensionless depth $\widetilde{\sigma}$. This calibration curve is shown in Fig.~\ref{fig:calibration}a. From the solution depicted in Fig.~\ref{fig:convolution}b another required calibration function is directly obtained to estimate ${\widetilde{\rho}}_t$ from $\widetilde{\sigma}$ at the reference intensity represented in the functional form as  
\begin{equation} \label{eq:calfun1}
{\widetilde{\rho}}_t=f_1(\widetilde{\sigma})
\end{equation}
\\
This calibration curve is illustrated in Fig.~\ref{fig:calibration}b. The input parameters for the analytical calibration functions $f_1$ and $f_2$ are conveniently measurable from the image. These functions can be further combined in the form ${\widetilde{\rho}}_t=f_1(f_2^{-1}(\widetilde{G}))$, as depicted in Fig.~\ref{fig:limits}a.

\begin{figure*}[ht]
\centering
\includegraphics[width=0.9\linewidth]{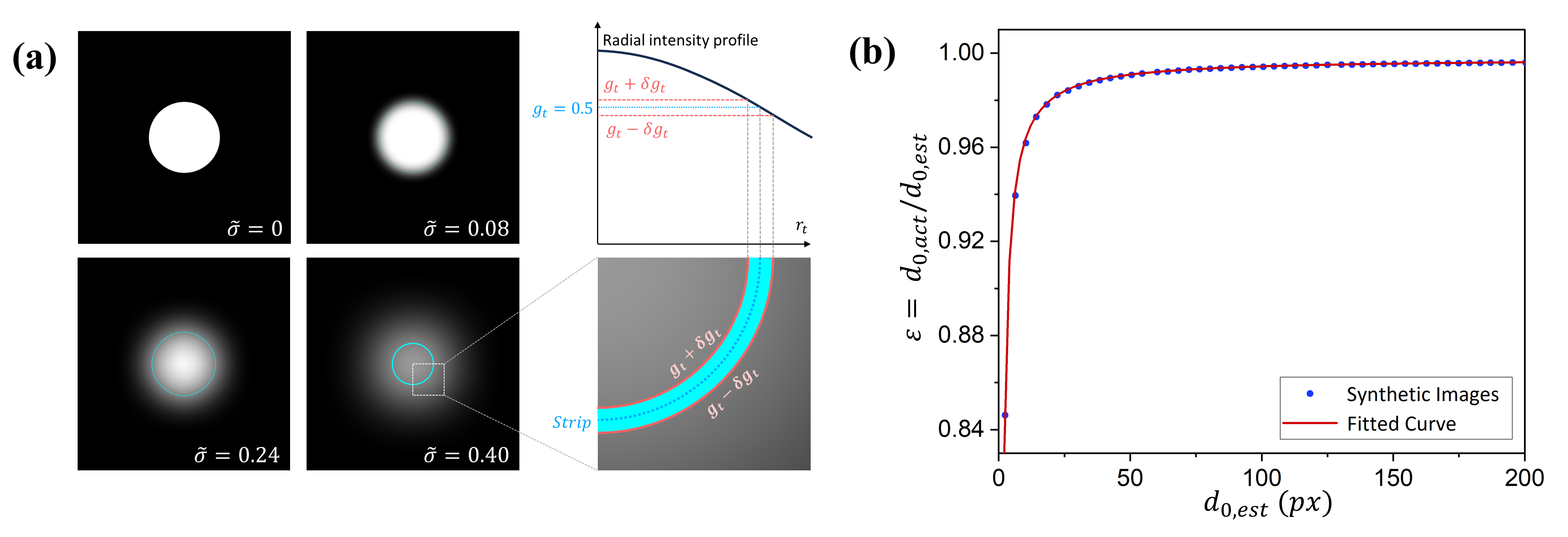}
\caption{(a) Gradient $\tilde{G}$ estimation using the average magnitude in a thin strip ($g_t \pm \delta g_t$) at the reference intensity depicted by cyan in the figure. This is necessary because the image is composed of pixels, restricting the precise estimation of gradients at exactly the reference intensity. The strip width increases as $\tilde{\sigma}$ increases, causing the average value to deviate from the anticipated exact value. (b) Error correction function $\varepsilon$ (ratio of actual to estimated diameter) generated using synthetic images to consider the pixelation effect on size and gradient estimation at reference intensity location $g_t$=0.5.}
\label{fig:error}
\end{figure*}

Being dimensionless, these analytical functions are universal to optical systems that exhibit a Gaussian blurring of the circular particles, which makes this technique a powerful measurement tool. The measurement process based on these functions is explained in the next subsection.
\\
\indent On the assessment of radial intensity profiles, i.e., $\rho_t-g_t$ curves  with varying $\widetilde{\sigma}$, the maximum slope values are found to occur at the intensity $g_t \approx 0.5$ for most of the suitable working range $\left(\widetilde{\sigma}\le0.2\right)$ (see Fig.~\ref{fig:limits}b). This intensity value at the location of maximum gradient magnitude, $g_t = 0.5$, is chosen as the reference location described earlier, making gradient estimation less susceptible to noise. The gradient $\widetilde{G}$ is estimated by considering the average magnitude within a thin strip whose edges are defined by the intensities $\left(g_t\pm\delta g_t\right)$ around the reference intensity (see Fig.~\ref{fig:error}a). This is necessary as the image is composed of pixels, and precise estimation at exactly the reference intensity is challenging. Moreover, the noise manifests as pixel level fluctuations, leading to sharp intensity variations; hence, steep local gradient values. By ensuring that the base gradient values are maximum at the region of interest, these fluctuations will have a minor influence on the estimated average when compared with the rest of the domain. 

The current analysis considers individual blurred particles, but in practical applications, particles often overlap when projected onto the image plane. This overlapping can result in a single indistinguishable, non-symmetric entity due to blurring. \textbf{Appendix~\ref{Appendix-A2}} includes a discussion on the particle concentration limit, which refers to the maximum degree to which closely packed particles can be distinguished. By solving the convolution equation specific to this case, it is deduced that the particles with a spacing between their centres greater than 1.4 times the diameter will be distinguishable at all depths for the segmentation threshold of 0.4. 

To the best of the authors' knowledge, this approach using both an intensity threshold and the gray level gradient for contour and size measurement is novel and a patent for this analytic approach has been filed.

\subsection{Measurement Process}
\vspace{3mm}
\textbf{Size estimation}: The size of the particles can be estimated based on the analytical calibration curves $f_1$  and $f_2$. First, the threshold radius $r_t$ and gradient magnitude $\left|\frac{\partial g_t}{\partial r_t}\right|$ is evaluated at the reference intensity  $g_t=0.5$ from the particle image. The associated image processing is explained in the \textbf{Materials and Methods} (Section~\ref{sec:Materials and Methods}), consisting of aspects like image normalization, segmentation, and sub-pixel interpolation. These  parameters are used to calculate the dimensionless 
gradient $\widetilde{G}=\left|\frac{\partial g_t}{\partial{\widetilde{R}}_t}\right|=\left|r_t\frac{\partial g_t}{\partial r_t}\right|$. From Eq.~(\ref{eq:calfun2}) the dimensionless depth $\widetilde{\sigma}=f_2^{-1}(\widetilde{G})$ is obtained and substituted into Eq.~(\ref{eq:calfun1}) to evaluate the dimensionless radius ${\widetilde{\rho}}_t=f_1\left(\widetilde{\sigma}\right)$. The  size of the particle in the image plane, $d_o$, is then evaluated using the relation $d_o\ ={r_t}/{{\widetilde{\rho}}_t}$.

\vspace{3mm}
\noindent \textbf{Depth estimation}: The estimation of particle depth requires an experimental calibration function in addition to the analytical functions used above. This step is optional and is not required if emphasis is placed only on the calibration free particle size estimation. Experimental calibration is achieved following the size estimation procedure described earlier and is performed for target dots or reticles of known size moved along the optical axis at known depths. The blur kernel size $\sigma$ is evaluated using the relation $\sigma= \widetilde{\sigma}d_o$. Since the depths of these target dots are already known, the correlation between $\sigma$ and $\left|\Delta z\right|$ can be estimated through Eq.~(\ref{eq:sig-delz}). The calculated linear fit $\beta$ remains constant for the system and is applied to the $\sigma$ values obtained from the sample particle measurements to estimate their corresponding depths. Due to the symmetric nature of the image blurring across the object plane, the depth location exhibits directional ambiguity, and only absolute values can be determined from the object plane.

Referring to Fig.~\ref{fig:limits}a, we now examine the characteristics of the calibration functions and their implications for the measurement process. In the vicinity of the object plane or the near-focus depth field  $\widetilde{\sigma}<0.05$, the parameter $\rho_t$ is practically constant, as can be seen in the combined calibration curve. This makes the method robust under near-focus conditions for diameter estimation, even though the gradient estimation and thus, $\widetilde{\sigma}$, is prone to error. This is due to the expected sharp gradients and the limitations imposed by image projection onto discrete pixels. Consequently, the depth estimates of particles near the object plane are unreliable. Furthermore, Fig.~\ref{fig:limits}a reveals a steep variation of ${\widetilde{\rho}}_t$ with the gradient in the blue shaded region corresponding to $\widetilde{\sigma}>0.35$. This region represents larger depth locations, approaching the limit of the measurement system. The measurements in this region are unreliable for diameter estimation. Moreover, the overall intensity level is lower due to a higher degree of blur, rendering the image susceptible to noise. This limits the measurement depth to approx. $\widetilde{\sigma}_c=0.35$, and the results beyond this are disregarded. Corresponding to this imposed limit $\widetilde{\rho_t}=0.3211$ and $\widetilde{G}=0.2501$. Hence, the availability of discrete two-dimensional intensity data due to pixelated image information poses a challenge in various ways. The errors associated with estimating gradients and threshold radius propagate through the aforementioned calibration functions, leading to inaccuracies in the estimated size values. To quantify this error, synthetic images of dots with known sizes and degrees of blur were analysed. An error correction function $\varepsilon$ is developed to compensate for the errors due to the pixelation effect, which is defined as the ratio:
\begin{equation}
\varepsilon(d_{0,est})=\frac{d_{0,act}}{d_{0,est}}
\end{equation}
\\
where $d_{0,est}$ is the diameter estimated using the proposed method and $d_{0,act}$ is the actual diamter of the particle. This function is illustrated in Fig.~\ref{fig:error}b and used to estimate the corrected diameter as $d_{0,corr} = d_{0,est} \cdot \varepsilon(d_{0,est})$. On closer inspection, we find the error in diameter estimation to be $\Delta d_{0} \approx 0.35$ pixel irrespective of the actual particle diameter for the proposed algorithm and parameters. Since the particle image is discrete, an inaccuracy of $\Delta d_{0} \approx 1$ pixel is anticipated; however, we are able to achieve a lower value due to the sub-pixel interpolation procedure discussed in the \textbf{Materials and Methods} (Section~\ref{sec:Materials and Methods}).

\subsection{Depth of Detection}
\vspace{3mm}

In the limit of detection corresponding to the depth $|\Delta z|=|\Delta z|_c$, the threshold radius $r_t \rightarrow 0$. Solving the dimensionless equation, Eq.~(\ref{eq:nondim-g_rt}) developed earlier,  this limit predicts a linear variation of depth of detection $\delta$ (total depth considering both sides of the object plane) with particle diameter $d_p$ \citep{sharma_depth_2023}, which can be represented as
\begin{equation} \label{eq:meas_depth}
    \delta = 2|\Delta z|_c = 2\alpha (d_p - d_{p0})
\end{equation}
\\
where $\alpha$ is a constant and $d_{p0}$ is an offset parameter to adjust the linear fit (usually $d_p \gg d_{p0}$). This offset parameter is an artefact of the pixelation associated with actual images and is discussed in detail in previous articles \citep{sharma_depth_2023}.

Considering the limit set on the measurement up to $\widetilde{\sigma}_c$, $\alpha$ can be determined using Eq.~(\ref{eq:sig-delz}), (\ref{eq:basic-nondim}) and (\ref{eq:meas_depth})
\begin{equation} \label{eq:alpha}
    \alpha=\frac{{\widetilde{\sigma}}_c}{\beta},\ \ {\widetilde{\sigma}}_c=0.35
\end{equation}
\\
The detection volume can then be determined as a function of particle size as
\begin{equation} \label{eq:meas_vol}
\begin{split}
    V_{d} & = \delta\left(H-d_{p}\right)\left(L-d_{p}\right) \\
    & = 2\alpha (d_p - d_{p0})(H-d_{p})(L-d_{p})
\end{split}
\end{equation}
\\
where $H\times L$ is the dimensions of the region of interest. This precise determination of the measurement volume is a distinguishing feature of the DFD approach, and a detailed discussion regarding the same can be found in previous works \citep{sharma_depth_2023}. The smaller particles are measured over a smaller depth range, and the detection depth increases linearly as the size increases, leading to an overweighting of larger particles. Hence, the information on detection depth is used for volumetric bias correction of the size distributions as discussed in the \textbf{Materials and Methods} (Section~\ref{sec:Materials and Methods}).

The parameter $\alpha$ plays a significant role in determining the detection volume ($V_d$), as indicated by Eq.~(\ref{eq:meas_vol}). This system parameter is dependent on $\beta$, implying that $\alpha \propto f/AD$ according to Eqs.~(\ref{eq:sig-delz}) and (\ref{eq:alpha}). Therefore, by choosing or adjusting these parameters, one can ensure a larger detection volume for a higher sampling rate. For instance, in designing the optical system for a particular application, if a larger focal length ($f$) for the optical system or or a lower aperture diameter ($D$) is chosen, one could achieve a larger detection volume.  Although the latter significantly affects the overall intensity profiles captured in the image and must be compensated by controlling the background illumination. Furthermore, experimental factors affecting parameter A are not precisely known, but it is highly dependent on the type, collimation, and chromaticity of background illumination. As will be demonstrated later using target dot measurements, a diffused beam illumination leads to a lower value of $\alpha$ and detection volume, but provides reliable measurement results. A collimated beam illumination however leads to a much higher $\alpha$ value, but the results obtained are unreliable due to the interference effects.

\section{Materials and Methods}
\label{sec:Materials and Methods}

\subsection{Experimental Setup}
\begin{figure} [ht]
\centering
\includegraphics[width=0.8\linewidth]{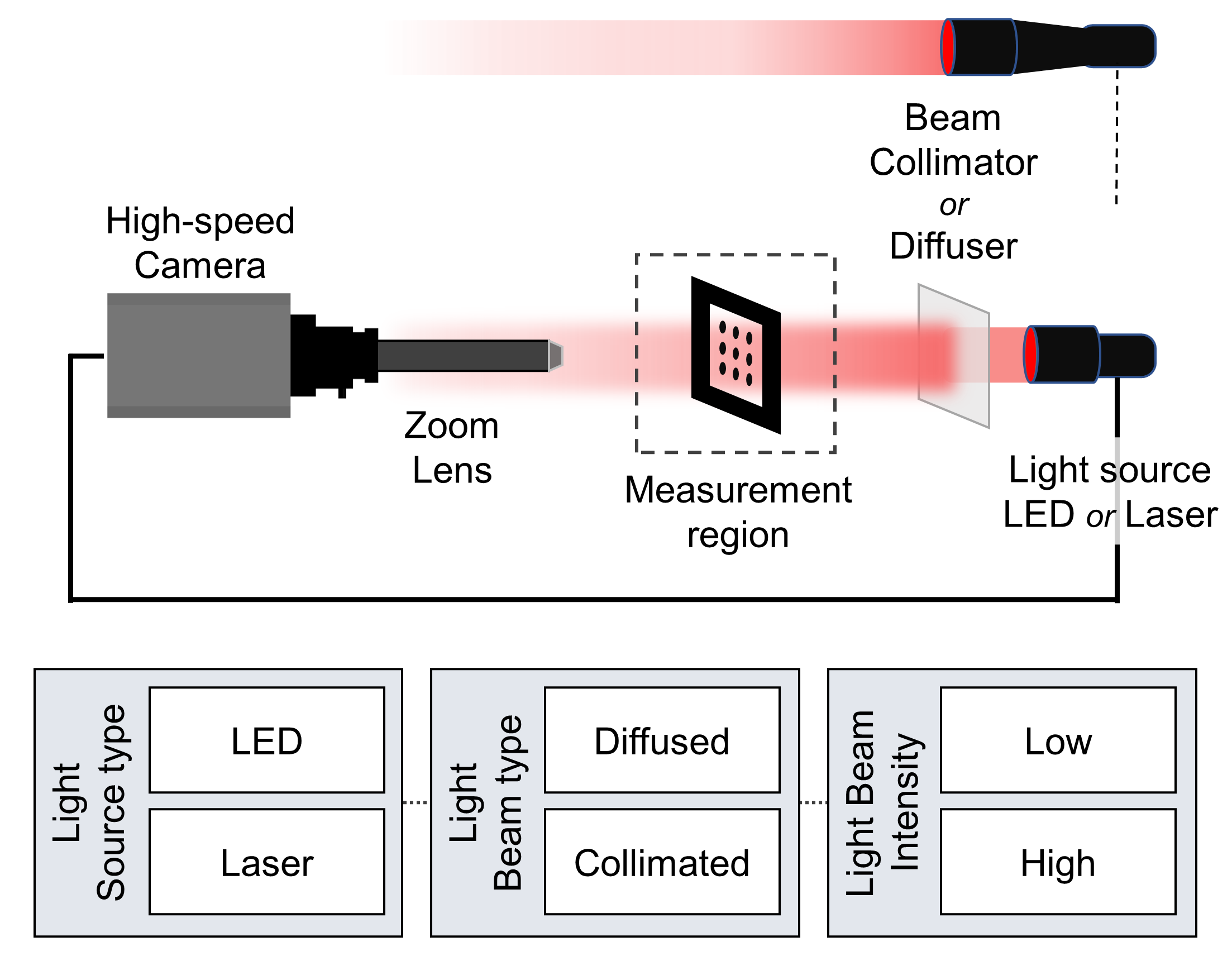}
\caption{Schematic of the optical arrangement for the single camera DFD measurement and various background illumination configurations to test the effect of parameters like chromaticity, collimation and intensity on measurements.}
\label{fig:setup}
\end{figure}

\textbf{Suitable Setup Requirements:}
\\
\noindent This measurement technique requires the minimal equipment associated with basic backlight imaging: a camera and a diffused light source for background illumination, as shown in Fig.~\ref{fig:setup}. For reliable measurements, the camera resolution and magnification should be carefully selected to ensure that the minimum particle of interest has a diameter of at least 3-5 pixels on the image sensor plane.
To achieve suitable background illumination, a diffusor plate or an appropriate optical device should be used. It is crucial to avoid collimated beams, as they can lead to inaccurate results due to non-Gaussian blurring and interference effects, such as Fresnel diffraction (refer to the \textbf{Appendix~\ref{Appendix-A1}}). Additionally, the light source should be aligned along the optical axis to ensure proper shadow formation, which means that the contours should remain circular when the particle is blurred. The background intensity should be adjusted to 
an intermediate value in the dynamic range of the sensor to avoid saturation associated with very high intensities and noise with low intensity levels. If the particle is not completely opaque, a bright central spot will appear inside the shadow, corresponding to first order refracted light passing through the particle. However, this effect can be more or less completely eliminated by moving the light source farther away from the object plane. In this manner, to an increasing degree, only paraxial rays will be seen and the intensity of the bright central spot decreases.  The formation of this localized central bright spot does not impact the estimation of radius and gradient at the reference intensity. 

The choice of lens is crucial and depends on the particle sizes being measured and the observation volume. A telecentric lens is preferred for accurate measurements, since it maintains a constant magnification, keeping the object size constant, independent of its position along the optical axis. Furthermore, a telecentric lens maintains symmetry of the blurred image for particles behind or in front of the object plane. However, standard optical arrangements can be used if the measurement volume is small in the depth direction, where the magnification variation is insignificant.
It is important to note that the aperture size and focal length can affect the system parameter $\beta$ (Eq.~(\ref{eq:sig-delz})) and consequently the measurement depth of the system (Eqs.~(\ref{eq:alpha}) and (\ref{eq:meas_vol})). 

\vspace{3mm}

\noindent \textbf{Setup used in Experiments}: 
\\
\noindent The basic configuration consisted of a high-speed camera, zoom lens and light source, with other accessories such as a beam expander, diffusor plate and calibration target dot plate.
\\
\noindent \textbf{Target Dot Measurement}:
High-speed camera: Photron SA5; 
Lens: 6.5$\times$ Navitar zoom lens coupled with 1.5$\times$ lens attachment, and 1$\times$ and 2$\times$ objective, where the latter was used for the higher magnification configuration; 
Light sources: Dolan Jenner Fiber-Lite Mi-150 LED light and Cavitar Cavilux smart UHS pulsed laser; 
Beam Expander: Thorlabs GBE05-A; 
Magnification: $\sim 6.8 \times$ and $\sim 13.7 \times$; 
Resolution: 2.94 $\mu$m/pixel and 1.46 $\mu$m/pixel.

\noindent \textbf{Glass Beads and Ethanol Spray Measurements}:
High-speed camera: Photron SA5; 
Lens: 6.5$\times$ Navitar zoom lens coupled with 1.5$\times$ lens attachment, and 1$\times$ objective; 
Light sources: Cavitar Cavilux smart UHS pulsed laser with diffusor plate; 
Magnification: $\sim 4 \times$; 
Resolution: 5 $\mu$m/pixel.

\noindent \textbf{Pollen Grain Measurement}:
High-speed camera: Photron SA5; 
Lens: 6.5$\times$ Navitar zoom lens coupled with 1.5$\times$ lens attachment, and 1$\times$ objective; 
Light sources: Dolan Jenner Fiber-Lite Mi-150 LED light with diffusor plate; 
Magnification: $\sim 4 \times$; 
Resolution: 5 $\mu$m/pixel.

\noindent \textbf{Bubble Rupture Aerosol Measurement and Surface Reconstruction}:
High-speed camera: Photron SA5; 
Lens: Tokina AT-X PRO M100 F2.8 D Macro lens; 
Light sources: Dolan Jenner Fiber-Lite Mi-150 LED light with diffusor plate; 
Magnification: $\sim 0.32 \times$; 
Resolution: 62.5 $\mu$m/pixel.

\subsection{Experimental Calibration Procedure}
The calibration procedure involves capturing a sequence of images to obtain the correlation between depth $|\Delta z|$ and blur kernel size $\sigma$. Hence, this step is optional and required only for depth estimation. We have confirmed a linear relationship between $\sigma$ and $|\Delta z|$ as depicted in the subsequent section in Fig.~\ref{fig:dots}c. The calibration target dots of known size are moved along the optical axis at known depth positions from the object plane. For each of these particles, blur kernel size or $\sigma$ can be estimated. Linear regression is performed on the scatter plot of $\sigma$ and $|\Delta z|$, as shown in Fig.~\ref{fig:dots}c, to derive the inverse functional form $|\Delta z| = m\sigma + c$. This functional form and associated parameters $(m,c)$ remain consistent for all measurements performed using the same optical system. Utilizing the calculated $\sigma$, along with the established functional form, we can estimate the depth of the particles under measurement. For improved accuracy, higher order polynomial fits can be considered.

If the object plane lies behind a glass window, then the calibration should ideally be conducted also with the glass window in place. The glass window will have the effect of shifting the absolute position measured by the system, but will not affect the relative positions between particles. If however, the dispersed phase is in a continuous phase with a refractive index other than air, then the value of $\beta$ will be affected. An example would be solid spheres in a liquid vessel, whereby the shadow imaging system is outside looking through the vessel. In this case, the calibration is best performed in situ, i.e., the calibration plate is traversed inside the vessel.

\subsection{Image Processing Algorithm}
The image processing routine consists of the following key aspects: Normalisation, Particle Identification, Sub-pixel interpolation, Size estimation and Depth estimation. The size and depth estimation processes utilise the proposed algorithm. The preceding steps are standard procedures for image processing systems. The flowchart for the algorithm depicted in Fig.~\ref{fig:flowchart} was implemented using MATLAB.

\noindent \textbf{Normalisation}: This process involves rescaling the intensity of the greyscale shadow image to a range of $[0,1]$. The global maximum value associated with the unobstructed illuminated background is mapped to 0, while the global minimum corresponding to the completely obstructed background or shadow is mapped to 1. The reference value for the former is derived from background illumination images and the latter from black-shading images (images captured with the camera lid on).  Mathematically, the normalised intensity $(I_n)$ is obtained \citep{zhou_spray_2020} as:
\begin{equation}
    I_n = \frac{I_{bi}-I}{I_{bi}-I_{bs}}
\end{equation}
where $I, I_{bi}$ and $I_{bs}$ are actual shadow image background illumination and blackshade image intensities, respectively.\\

\begin{figure*}[ht]
\centering
\includegraphics[width=0.65\linewidth]{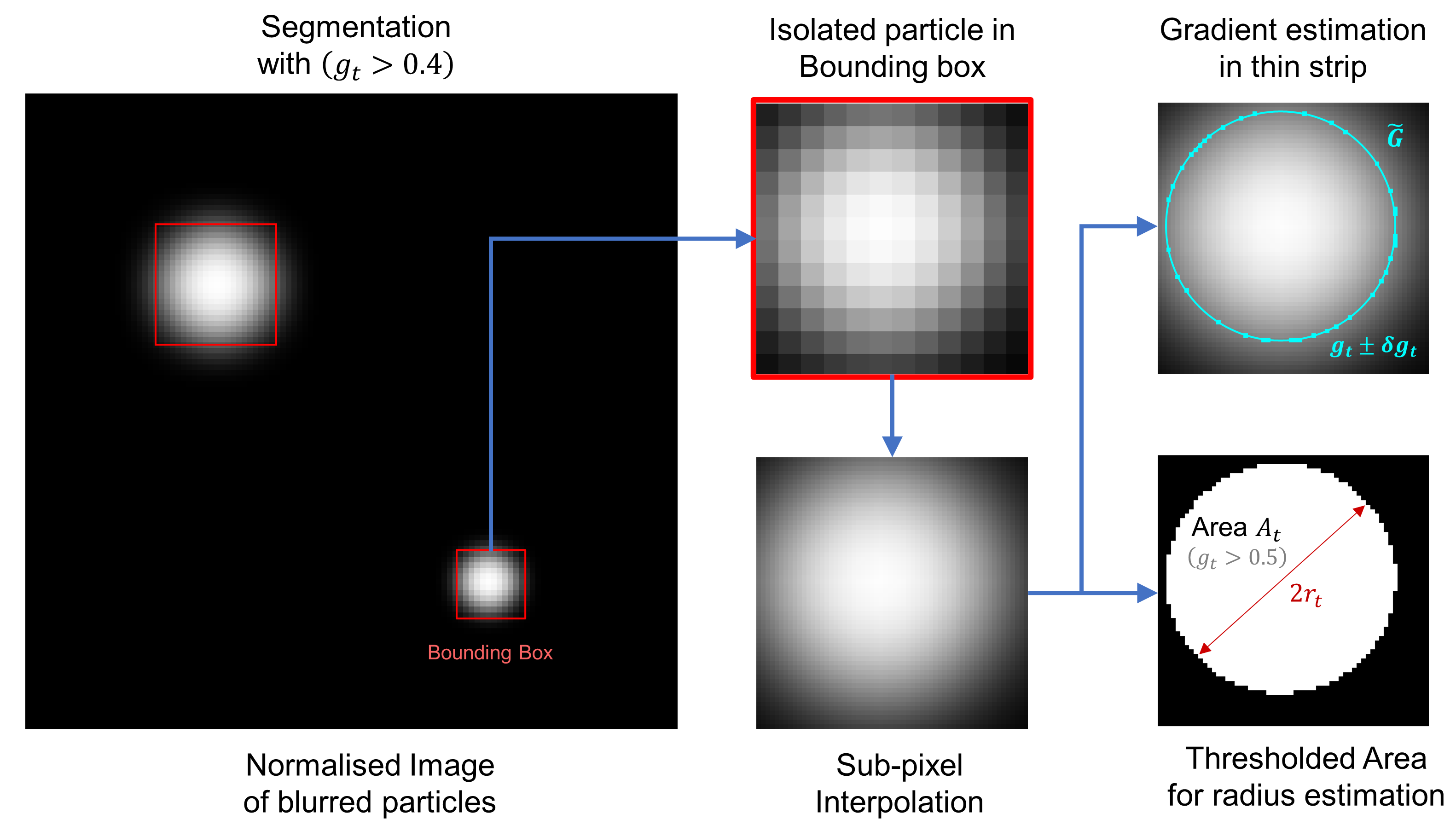}
\caption{Image processing steps depicting segmentation of normalised image and extracting image of each particle enclosed in a bounding box, sub-pixel interpolation, thresholding to estimate radius $r_t$ and average gradient $\widetilde{G}$ within a thin strip defined by edges at ($g_t \pm \delta g_t$)}
\label{fig:processing}
\end{figure*}

\noindent \textbf{Particle Identification}: This step involves isolating and extracting individual particles from the normalized image for further analysis. In this study, a simple intensity based method was adopted, in which regions with an intensity above a threshold value were identified as a particle. This process, known as segmentation in image processing, allowed for particle identification with a threshold set at 0.4 for this study. The particles were isolated as separate images based on the bounding box enclosing the identified regions on segmentation (see Fig.~\ref{fig:processing}). The bounding box refers to the smallest rectangular region that encloses the particle. The intensity threshold for particle detection should be lower than the reference intensity value of 0.5, within which the subsequent analysis for size and depth estimation is conducted. This ensures that the information used for estimation is extracted within the bounding box, sufficiently away from its edges. Depending on the system under study, more advanced algorithms can be employed for the segmentation or isolation process.

\noindent \textbf{Sub-pixel Interpolation}: This step involves interpolating intensity data on a grid finer than pixel resolution for the isolated particles. This is necessary because only discrete information is available from an image, and extraction of information precisely at exactly some prescribed reference intensity is a challenge. In this study, a simple bilinear interpolation was performed, where each pixel was subdivided into a $5\times 5$ grid (see Fig.~\ref{fig:processing}). Prior to the interpolation process, a noise removal step is performed using a Wiener filter. Depending on the noise characteristics of the system, further advanced interpolation techniques can be performed on a suitable sub-grid. 

\noindent \textbf{Size Estimation}: To estimate the image size $d_o$ of the isolated particle, radius and gradient magnitude information at a reference intensity of 0.5 is required. The radius $r_t$ is determined by obtaining a region with an intensity above 0.5 and calculating the equivalent radius from its area $A_t$ as $r_t = \sqrt{A_t / \pi}$ (see Fig.~\ref{fig:processing}). If glare points exist, they will appear as holes in this image region and can be easily removed by the 'fill hole' operation commonly available in image processing systems. The region eccentricity provides an estimation of the actual particle shape and is used to segregate non-circular particles as discussed in the subsequent sections. To determine the gradient, the average magnitude in a thin strip ($g_t \pm \delta g_t$) centred at reference intensity is considered (Fig.~\ref{fig:error}a and Fig.~\ref{fig:processing}). The gradient can be calculated using standard gradient functions available in image processing systems. For this study, the strip width is set by choosing $\delta g_t = 0.005$. 

The threshold radius $r_t$ and gradient magnitude $\left|\frac{\partial g_t}{\partial r_t}\right|$ evaluated as above are then used to determine the dimensionless gradient $\widetilde{G}=\left|r_t\frac{\partial g_t}{\partial r_t}\right|$. The analytical calibration functions $f_1$ and $f_2$ are employed to determine $\widetilde{\sigma}=f_2^{-1}(\widetilde{G})$ and subsequently, ${\widetilde{\rho}}_t=f_1\left(\widetilde{\sigma}\right)$. The size of the particle $d_o$ is evaluated using the relation $d_o\ ={r_t}/{{\widetilde{\rho}}_t}$. The blur kernel size $\sigma$ is evaluated using the relation $\sigma= \widetilde{\sigma}d_o$. Until this step, analytical functions are sufficient and experimental calibration is not required. Hence, size estimation can be performed independently in a calibration free manner.

\noindent \textbf{Depth Estimation}: To estimate depth, the inverse functional form $|\Delta z| = m\sigma + c$ from the experimental calibration procedure is required. By substituting the determined value of $\sigma$, the absolute depth from the object plane is evaluated. However, the proposed method does not provide directional information for the depth.

\subsection{Limiting Parameters for Reliable Measurements}
Particles located at the outer limits of the detection depth exhibit high levels of blurring, low intensities, and significant alterations in gradients due to imaging system noise. Consequently, measurements in this region are highly unreliable, as even a small error in gradient estimation can result in a large diameter error. To address this, we introduce a critical measurement depth limit $\widetilde{\sigma}_c=0.35$, beyond which results are not considered. By imposing a tighter depth of detection with a lower $\widetilde{\sigma}_c$ value, more accurate overall results can be achieved. Furthermore, while the ideal eccentricity for spherical entities is zero, a practical limit can be set in the range of 0.5 to 0.8. Particles exceeding this limit can be rejected from the analysis.

\subsection{Volumetric Corrections in Size Distributions}
The detection depth and volume are dependent on the size of the particle being measured. Detection depth varies linearly with particle size, and the detection volume can be determined as per Eq.~(\ref{eq:meas_vol}). This leads to a volumetric measurement bias, because larger particles are measured (and counted) over a larger volume compared to the smaller particles. To address this bias, it is important to consider the number of dispersed particles per unit volume when determining the size distribution. This can be achieved by weighting the occurrence frequency in each histogram bin by the inverse of the corresponding measurement volume. Normalizing this weighted frequency yields the required probability density function. From Eq.~(\ref{eq:meas_vol}) it can be observed that $d_{p0}$ is not significant, since $d_p \gg d_{p0}$ and $V_d \propto \alpha$, 
 which implies that this $\alpha$ will cancel out uniformly during the normalisation procedure. Hence, the volumetric bias correction of the PDFs can be easily achieved without any experimental calibration or  knowledge of $\alpha$. This estimation of the size probability density distribution implicitly assumes that that the distribution is uniform along the optical axis. Nevertheless, since the position and size of all particles are known, one could retroactively examine subvolumes and determine whether the assumption of uniformity was correct. However, the subvolumes must lie within the detection bounds of all particles.

\section{Results}
\label{sec:Results}

\subsection{Parameteric analysis of measurement system}
\begin{figure*}[ht]
\centering
\includegraphics[width=\linewidth]{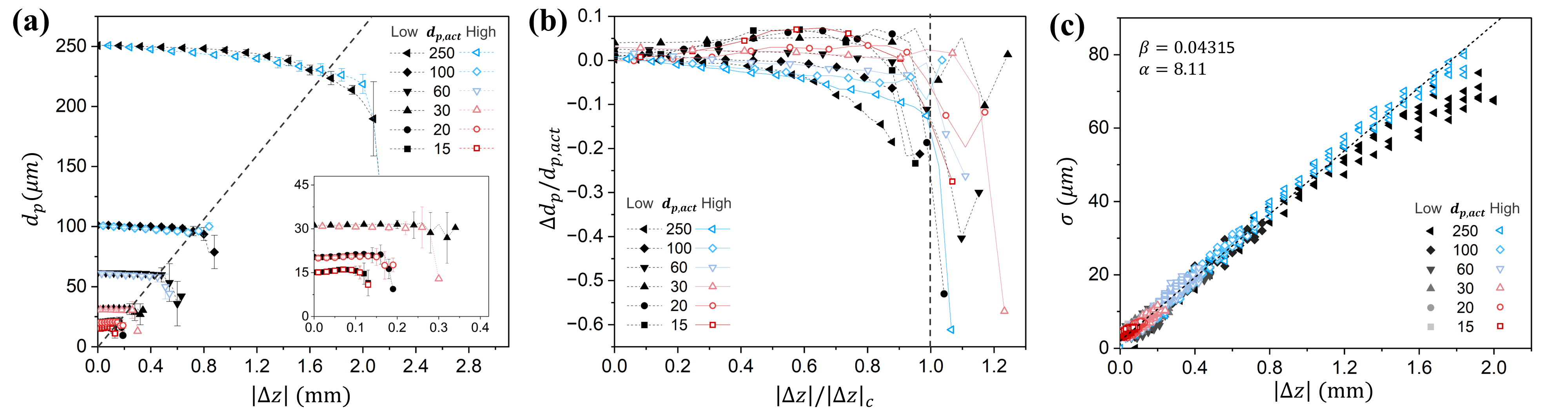}
\caption{Measurement results for calibration dots of known sizes and depths at a magnification $\sim$6.8x for diffused LED beam illumination depicting the variation of (a) measured diameter with depth (b) the relative error in diameter measurement with dimensionless depth (c) blur kernel size with depth. ‘Low’ and ‘High’ intensity background illumination measurements are overlaid on the same plot.}
\label{fig:dots}
\end{figure*}

\vspace{3mm}
The calibration target dots (or reticles) of known size are moved along the optical axis at known depths and captured in different background illumination configurations (see Fig.~\ref{fig:setup}). This enables to validate the measurement technique by comparing the size estimated by the proposed technique with the actual dot size at various depth locations. A comprehensive discussion on various illumination configurations using diffused and collimated light can be found in the \textbf{Appendix~\ref{Appendix-A1}}. Measurements for the case of a diffused LED light source illumination are performed at a magnification of $\sim6.8$x at two background intensity levels (low(0.2) and high(0.65), rescaled average background image pixel bit value where 0.2 means intensity at $20\%$ of the dynamic range of the image sensor where $100\%$ represents completely saturated) and depicted in Fig.~\ref{fig:dots}. The size is predicted accurately up to a 5-15\% relative error in most parts of the measurement depth (Fig.~\ref{fig:dots}b). One observes a higher relative error in measurements for the collimated beam illumination due to the interference pattern caused by Fresnel diffraction \citep{hecht_optics_2012}. Hence the proposed analysis does not apply to such optical settings due to the non-Gaussian blurring \citep{stokseth_properties_1969, lee_review_1990} of the dots. The dashed line in Fig.~\ref{fig:dots}a represents the linear depth of detection, indicated by $\widetilde{\sigma}_c=0.35$. Measurements beyond this limit on the right side are not as unreliable as anticipated. The target dots measurements also enable to validate the hypothesis of a linear relationship $\sigma \propto |\Delta z|$, as depicted in Fig.~\ref{fig:dots}c. Hence, the experimental calibration can be performed and $\beta$ can be estimated through linear regression from these dot images. No considerable effect of the background illumination intensity is observed. Still, 
an intermediate background intensity is suggested, as a lower value is prone to noise, and a higher value might flush out the blurring information due to over-saturation at the sensor. For particles of the same physical size, the higher magnification ensures the availability of more pixels to extract more accurate information. This enables a slightly better estimation of size. A discussion on measurements at higher magnification ($\sim13.7$x) is presented in the \textbf{Appendix~\ref{Appendix-A1}}.

\subsection{Technique implementation for diverse applications}
This section illustrates the application of the technique to a diverse range of problems. The details of the experimental setup for each system are provided in \textbf{Materials and Methods} (Section~\ref{sec:Materials and Methods}).

\subsubsection{Dispersed Glass beads}
\begin{figure}[ht]
\centering
\includegraphics[width=1\linewidth]{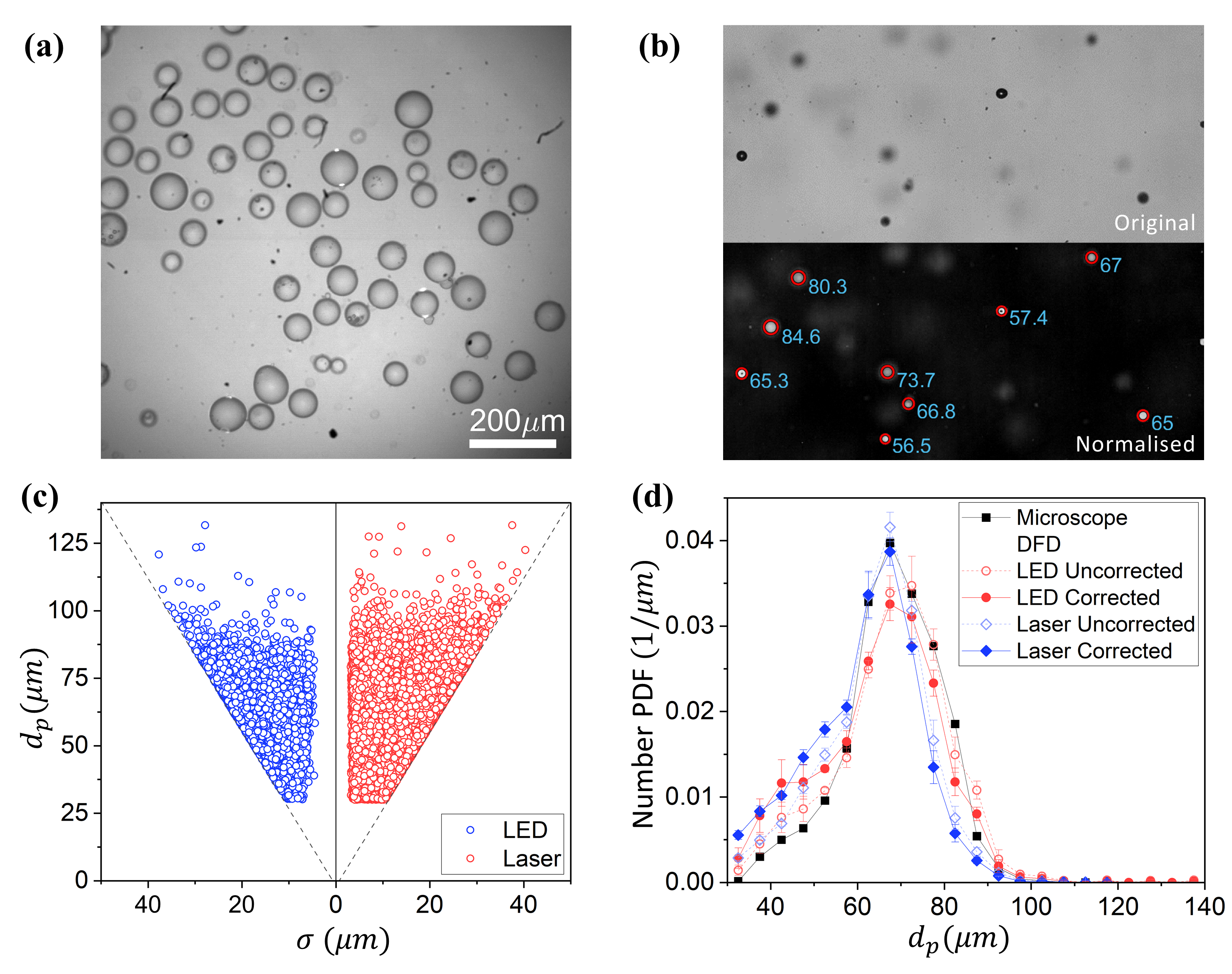}
\caption{Measurement results for diverse applications utilising a diffused background illumination (a) Spherical glass beads under the microscope (b) Glass beads dispersed in a solution, being continuously stirred. The detected beads are marked with red circles in the normalised shadow image with size $d_p$ in $\mu$m (c) The estimated size of dispersed glass beads $d_p$ and the corresponding blur kernel size $\sigma$ depicting the linear relationship between the depth of detection and the diameter. (d) Comparison of the size distribution evaluated from the DFD technique with the microscope measurements as a reference. The uncorrected and detection volume bias-corrected estimates are depicted as Probability Density Functions (PDFs).}
\label{fig:beads}
\end{figure}

Untinted spherical glass beads within a size range of $40-90 \mu m$ are used for sample measurement. Such measurements are common in the field of chemical sciences, particularly as calibration standards for a wide range of analytical techniques such as flow cytometry and spectroscopy. To validate the approach, a reference size distribution is estimated using microscope images of glass beads on a slide (Fig.~\ref{fig:beads}a). For measuring in a DFD system, the glass beads are uniformly dispersed in a DI water solution and stirred continuously to avoid settling. Shadow images of the dispersed solution are captured using a  diffused LED and laser illumination (Fig.~\ref{fig:beads}b). The predicted size distribution from the DFD measurement is compared with the microscope results and is in good agreement (see Fig.~\ref{fig:beads}d). Error bars are added to represent one standard deviation realised over six runs. The volumetric measurement with a varying detection depth is evident from Fig.~\ref{fig:beads}c. 

\subsubsection{Sprays}
The measurement of a droplet size distribution in sprays holds significance in various natural and industrial systems. For instance, in fuel injection systems, the size of atomized droplets affects combustion efficiency through droplet lifetime and evaporation rate \citep{kumar_experimental_2022}. In high-speed gas flow-induced atomization, precise control of droplet dispersion size is important for monodisperse powder production for additive manufacturing and pharmaceutical applications \citep{sharma_shock_2021, sharma_advances_2022}. The COVID-19 pandemic highlighted the role of micro-droplets in disease transmission and the requirement to develop mitigation strategies \citep{fischer_low-cost_2020,prather2020airborne,sharma_secondarymask_2021}. To illustrate the applicability of the DFD method, shadow imaging of an ethanol spray using monochromatic background illumination from a diffused laser beam is performed (Fig.~\ref{fig:spray}a). The spray is generated using a laboratory grade positive displacement pump-type spray dispenser. Measurements are performed at a downstream sparse spray region to obtain the size distribution, as depicted in Fig.~\ref{fig:spray}b. The error bars correspond to the standard deviation evaluated from six runs. The number distribution follows a familiar skewed distribution, commonly observed in dispersed spray systems.

\subsubsection{Aerosol generation from surface bubble rupture}
\begin{figure}[ht]
\centering
\includegraphics[width=1\linewidth]{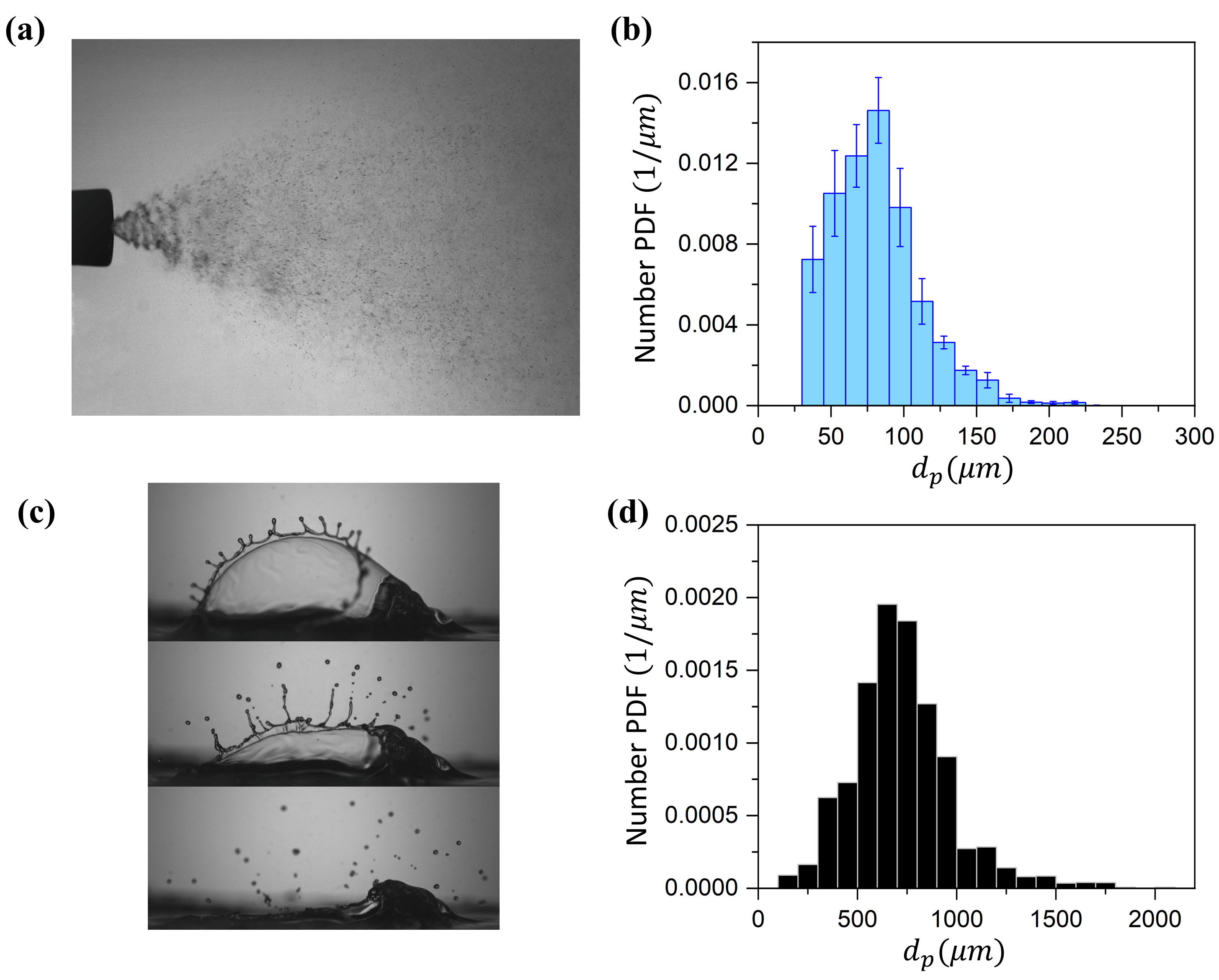}
\caption{(a) Ethanol spray in monochromatic background, illumination using a diffused laser beam. (b) Ethanol spray droplet size distribution measured using the DFD technique, represented as a PDF corrected for detection volume bias. (c) Droplets generated during the rupture of a surface bubble in DI water. (d) Droplet size distribution from bubble rupture measured using the DFD technique represented as corrected PDF.}
\label{fig:spray}
\end{figure}

Air bubbles formed at a liquid surface undergo film drainage, eventually leading to rupture and fragmentation into dispersed droplets (Fig.~\ref{fig:spray}c). Depending on the surface tension, film thickness, and the bubble lifetime, this can lead to the formation of droplets in the aerosolization range \citep{lhuissier_bursting_2012}. This mode of mass transfer at bulk liquid interfaces is of interest in marine and environmental sciences. Furthermore, recent studies \citep{poulain_biosurfactants_2018} identified the effect of biological secretions on the size of fragmenting droplets, with many falling in sizes critical for aerosolization. Such transport of pathogen-loaded droplets into the ambient environment is relevant to disease transmission. Hence the proposed method can be deployed for such studies.
To illustrate this method, bubbles are generated below the surface of a sample liquid pool with a nozzle connected to the air supply from the pump. The continuous bubbles generated in the DI water sample coalesce to form a larger surface bubble of diameter $\sim$30 mm (spherical cap), which eventually ruptures. For measurement, shadow imaging is performed on the unobstructed dispersed droplets generated from the rupture of a bubble, and $\sim$50 such events were considered. The obtained size distribution is depicted in Fig.~\ref{fig:spray}d.

\subsubsection{Pollen viability}
The  health of pollen grains regarding their ability for germination depends on numerous factors, including their diameter and shape \citep{kelly_method_2002}. In some species, the inviable or unhealthy pollens are smaller and oddly shaped, associated with aberrant or dehydrated conditions. Pollen grains have a wide range of shapes, with sizes ranging from 10-200 microns. There are a limited number of methods available for their size characterisation \citep{de_storme_volume-based_2013, kumar_quantification_2022}, and the newly proposed method in this study can serve as a simple tool to characterize near-spherical pollens dispersed in a solution. To illustrate this, a pollen sample of Hibiscus (\textit{Hibiscus rosa-sinensis}, see Fig.~\ref{fig:misc}a) is collected from the institute gardens and dispersed in DI water. Hibiscus has spherical shaped pollen grains in the size range of 80-180 $\mu m$, with very small spike like features on its periphery \citep{shaheen_pollen_2009}. The size distribution obtained by implementing the DFD technique on shadow images is depicted in Fig.~\ref{fig:misc}b. The error bars correspond to the standard deviation realised over five runs.

\subsubsection{Surface reconstruction}
Digital reconstruction of a three-dimensional surface is an important application in computer vision \citep{ghita_video-rate_2001, saxena_3-d_2008} or interfacial fluid mechanics \citep{willert_three-dimensional_1992}. The proposed method can be implemented in this context by considering patterned surfaces, i.e., engraving target dots over the surface of interest and performing the analysis to obtain the depth profiles. As the blurring is still Gaussian, the same theory and calibration functions are applicable. To illustrate this, images of patterned surfaces with target dots printed on paper adhering to a defined three-dimensional (3D) contour are considered.

\begin{figure}[ht]
\centering
\includegraphics[width=1\linewidth]{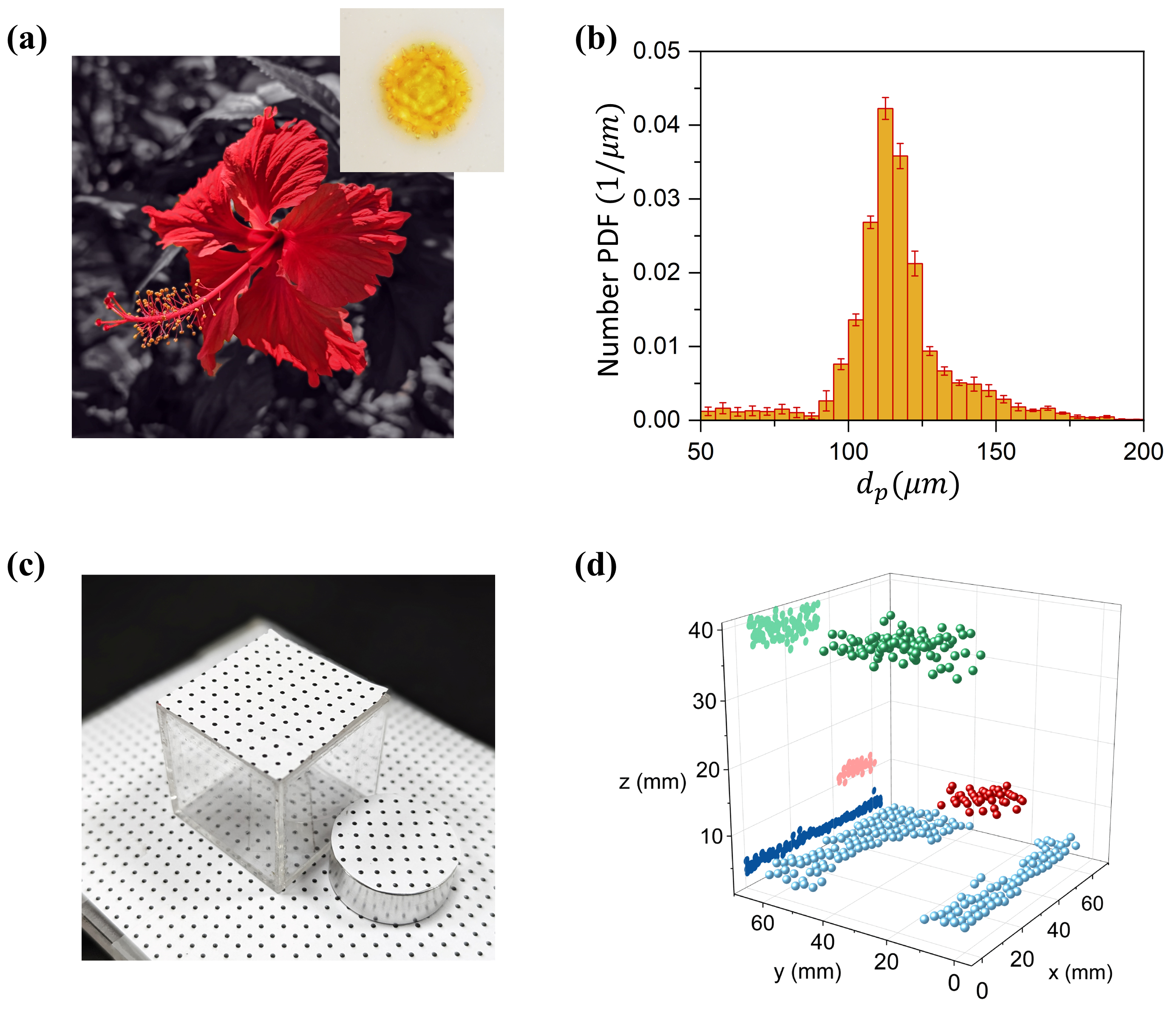}
\caption{(a) Hibiscus (\textit{Hibiscus rosa-sinensis}) flower and its pollen grain under the microscope. (b) Pollen grain size distribution estimated using the DFD technique represented as a corrected PDF. (c) Target dots engraved over a prescribed 3D surface (d) Digital reconstruction of this 3D surface geometry using the DFD technique with surface points projected on the x-z plane to depict depth. Error bars, when represented, correspond to one standard deviation in the size distributions.}
\label{fig:misc}
\end{figure}

A camera with normal front lighting from an LED source is chosen, i.e., not shadow imaging, to depict the simplicity and versatility of this method. The 3D geometry and resultant scanned surface are depicted in Fig.~\ref{fig:misc}c,d. The actual height of the cuboidal (green) and cylindrical (red) surfaces was 33mm and 10mm, which is measured to be approximately 34mm and 8mm, respectively. The reconstructed profiles closely matched the actual shape, but discrepancies were observed for the surface near the focal plane. This is due to the inaccurate estimation of sharp gradients from discrete pixel information. Improved depth estimations can be achieved by keeping particles or surface features within the intermediate regions of the depth of detection. Hence, a simple extension of the proposed algorithm can be used for the digital reconstruction of patterned 3D surfaces. As the method estimates the degree of blur of the scene, this can be reinforced with deconvolution algorithms as well for deblurring operations. In the context of experimental fluid mechanics, this method is applicable for reconstructing a free surface of the fluid with floating spherical particles that serve as target markers.

\section{Discussion}
\label{sec:Discussion}
We introduce a new measurement technique to precisely characterise the size and position of both in-focus and out-of-focus spherical dispersions using minimal and accessible optical resources. The measurement principle is based on an analytical framework of image blurring, and the derived functions are universal, enabling particle sizing in a calibration-free manner. Particle position from the object plane is estimated based on its correlation with the degree of blurring, established using a simple calibration procedure. The system precisely calculates measurement volume and its dependence on the size of the dispersion particles. This is crucial to obtain bias-free size distribution and volume concentration estimates. The method requires simple shadow imaging with a diffused light source for background illumination and a camera, paired with a telecentric lens or equivalent arrangement. With a suitable spatiotemporal resolution, implementation is possible in various systems, including microns to millimetres size particles moving with speeds ranging from stationary suspensions to supersonic droplets, limited only by imaging hardware capabilities.

To validate the method, opaque target dots of known size at known incremental depth locations across the object plane were considered. The implementation under various background illumination demonstrated its suitability in diffused beams, where the blurring is Gaussian. However, in cases with collimated beams, the presence of diffraction effects resulted in deviations due to the non-Gaussian nature of the point spread function (PSF), in particular for very small particles.
\\
To illustrate the technique, sparse dispersions of spherical particles like glass beads, spray droplets and pollen grains were considered. In the case of dispersed glass beads, microscopy was used as a reference to validate the DFD measurements. The technique was further extended to computer vision applications, where a three-dimensional surface profile was reconstructed digitally using engraved target dots. The resultant profile matched the actual shape, although discrepancies were observed for surfaces near the focal plane due to the inaccurate estimation of sharp gradients.
\\
It should be noted that the measurement accuracy is limited by the precision of gradient evaluation from discrete pixel information, which is susceptible to noise. Moreover, although the absolute distance of the particle from the object plane is known, an ambiguity remains whether the particle is positioned in front of or behind the object plane. Thus, in practice the optical arrangement should be designed, such that the region of interest lies all on one side of the object plane, to avoid ambiguous position measurements. Note that this ambiguity does not exist for the two-camera implementation of the DFD technique.

The question may be posed whether a position measurement of each particle is necessary, since this requires the extra calibration step? There are several reasons why this might be essential. For one, if particle tracking is to be realized, for instance with a high-speed camera, then the particle position must be known at each time step. The position would also be necessary if spatial inhomogenieties of size distribution are to be detected.

As an outlook, the approach using blur gradients together with a gray level threshold offers possibilities in characterising overlapping projections in dense particle clusters and/or non-spherical/irregular particles. The firsts extension would greatly increase the tolerable volume concentration for applying this technique. The second feature would open up inumerable new application areas. Both of these extensions are currently being developed by the authors.


\begin{appendices}

\section{Parameteric analysis of measurement system}\label{Appendix-A1}
\begin{figure*}[ht]
\centering
\includegraphics[width=0.75\linewidth]{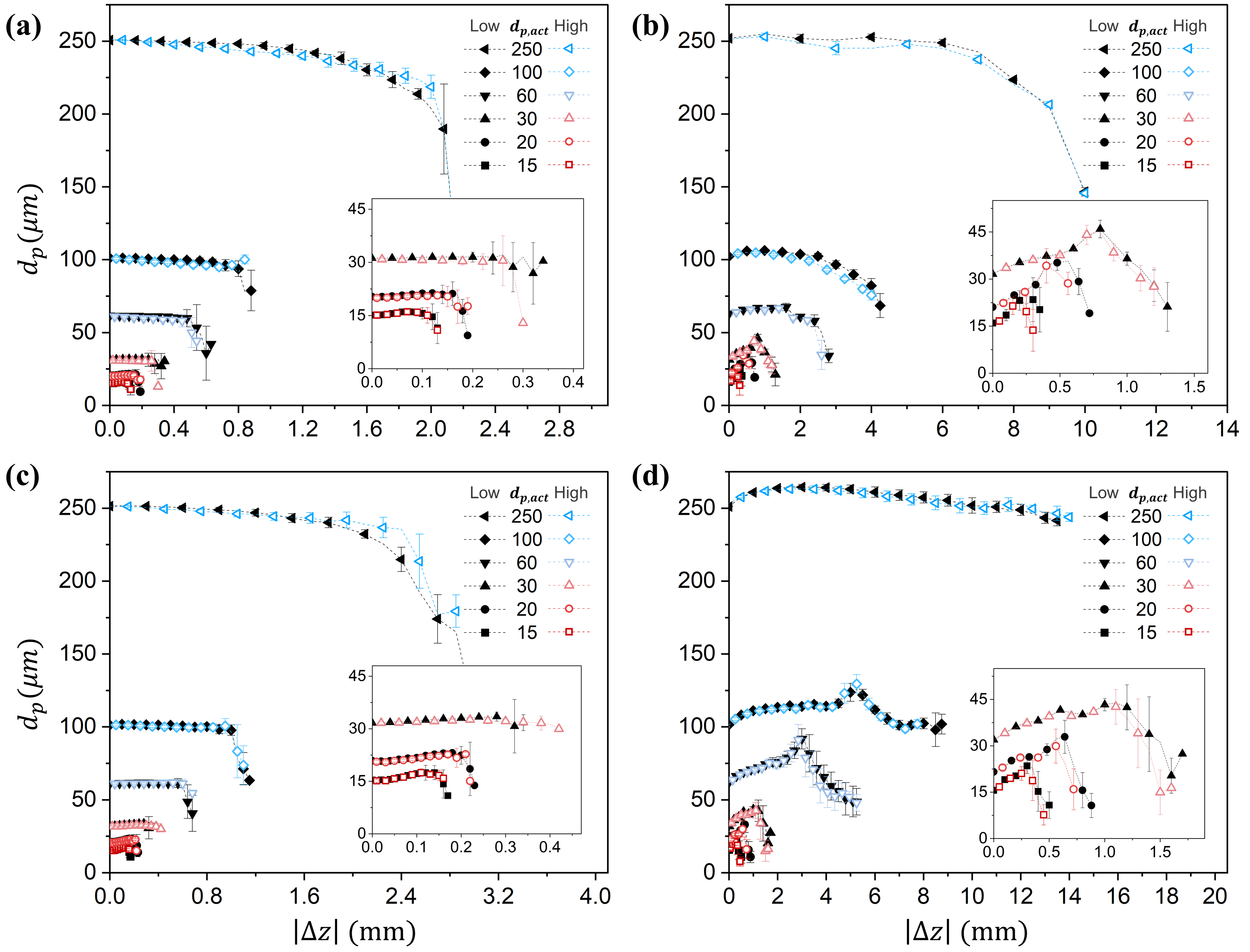}
\caption{Measurement results for calibration dots of known size and depth illustrating the variation of measured diameter with depth for different background illumination configurations (a) LED white light diffused beam (b) LED white light collimated beam (c) Laser mono-chromatic light diffused beam (d) Laser monochromatic light collimated beam. ‘Low’ and ‘High’ intensity illumination measurements are overlaid on the same plot for magnification $\sim$6.8x.}
\label{fig:figures1}
\end{figure*}

\begin{figure*}[ht]
\centering
\includegraphics[width=0.75\linewidth]{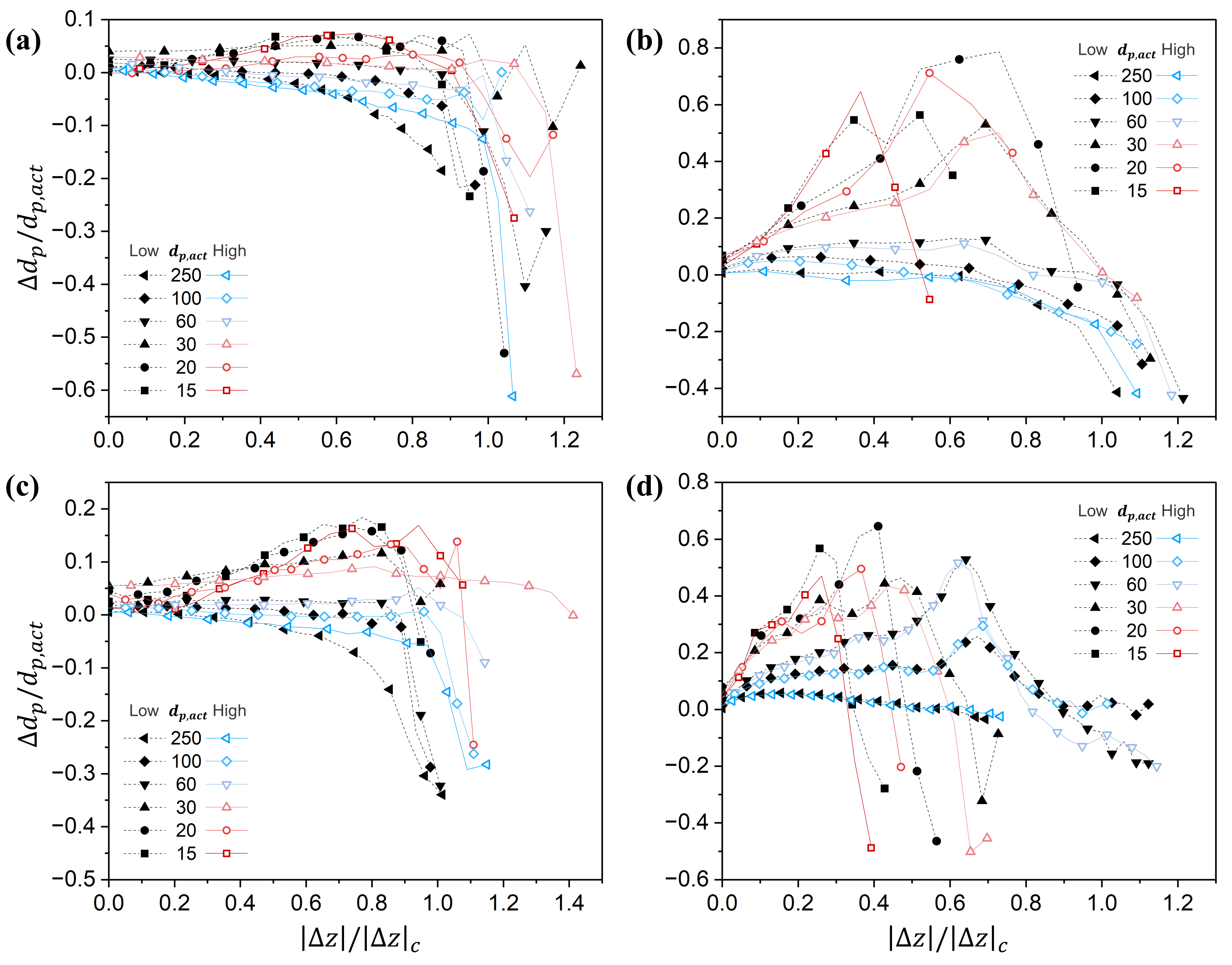}
\caption{Measurement results for calibration dots of known size and depth illustrating the variation of relative error in measured diameter with dimensionless depth for different background illumination configurations (a) LED white light diffused beam (b) LED white light collimated beam (c) Laser mono-chromatic light diffused beam (d) Laser monochromatic light collimated beam. ‘Low’ and ‘High’ intensity illumination measurements are overlaid on the same plot for magnification $\sim$6.8x.}
\label{fig:figures2}
\end{figure*}

\begin{figure*}[ht]
\centering
\includegraphics[width=0.75\linewidth]{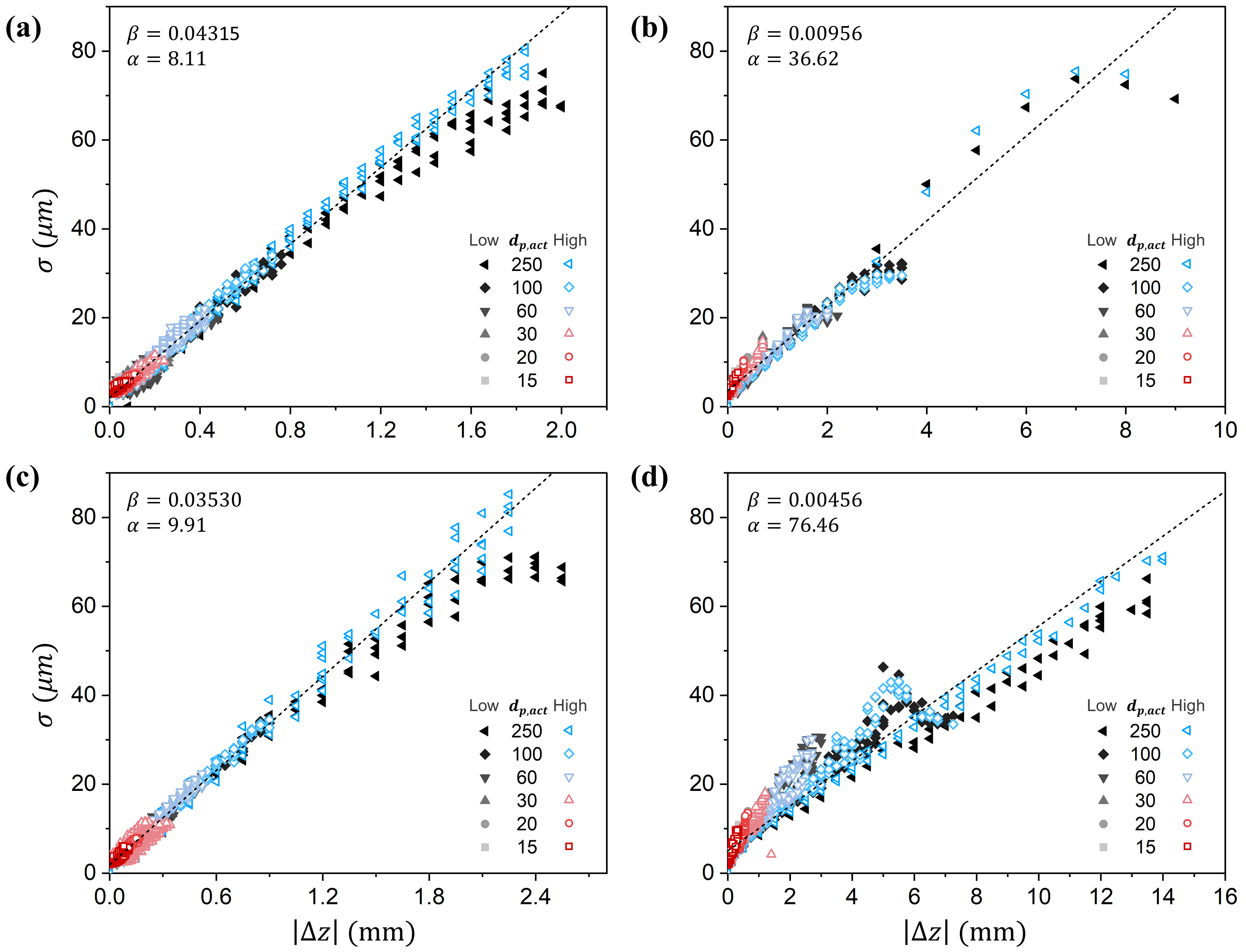}
\caption{Measurement results for calibration dots of known size and depth illustrating the variation of blur kernel size with depth for different background illumination configurations (a) LED white light diffused beam (b) LED white light collimated beam (c) Laser mono-chromatic light diffused beam (d) Laser monochromatic light collimated beam. ‘Low’ and ‘High’ intensity illumination measurements are overlaid on the same plot for magnification $\sim$6.8x.}
\label{fig:figures3}
\end{figure*}

\begin{figure*}[ht]
\centering
\includegraphics[width=\linewidth]{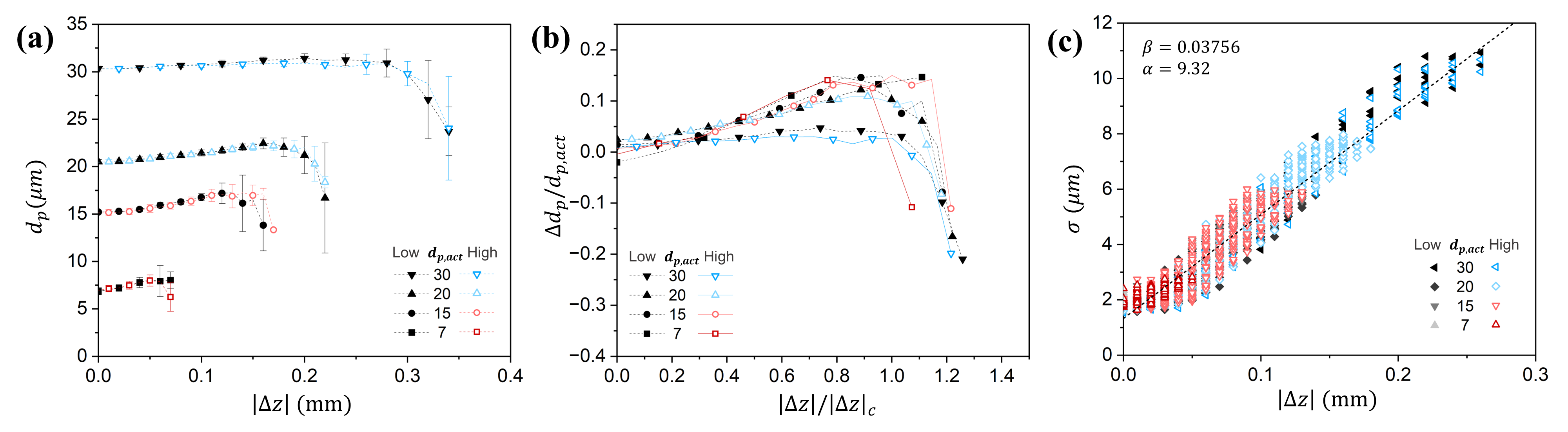}
\caption{Measurement results for calibration dots of known size and depth at a higher magnification $\sim$13.7x for monochromatic diffused laser beam illumination illustrating the variation of (a) measured diameter with depth (b) the relative error in measured diameter with dimensionless depth (c) blur kernel size with depth. ‘Low’ and ‘High’ intensity illumination measurements are overlaid on the same plot.}
\label{fig:figures4}
\end{figure*}

\begin{figure*}[ht]
\centering
\includegraphics[width=0.6\linewidth]{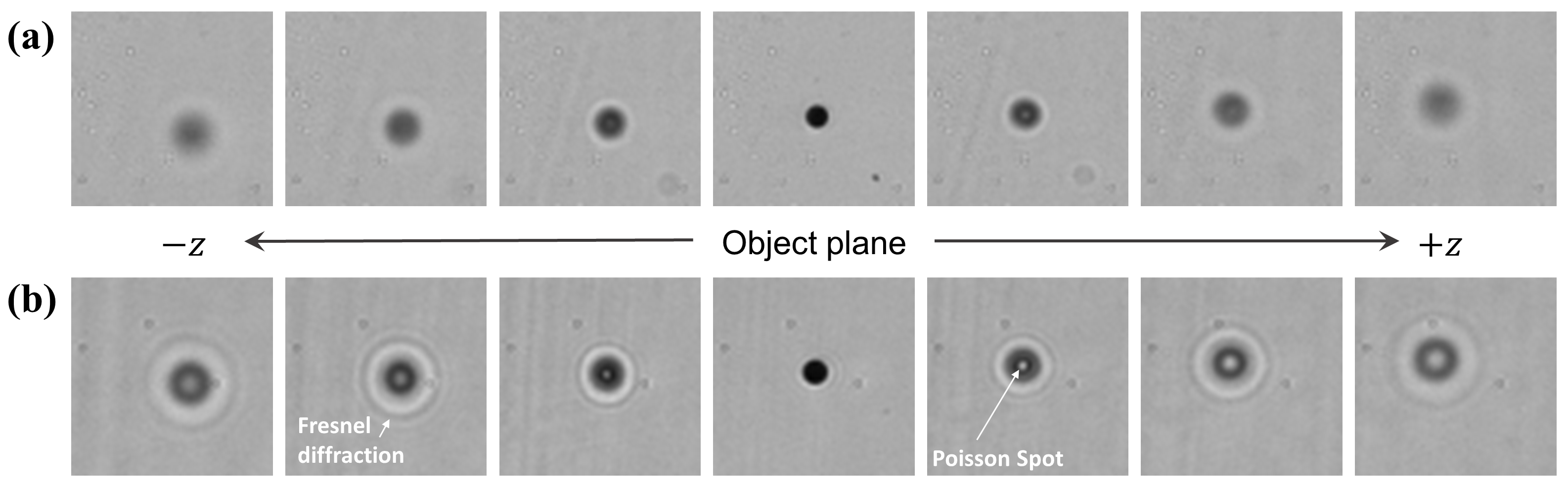}
\caption{Non-Gaussian blurring of a calibration target dot (diameter-20 $\mu$m) in collimated background illumination (a) LED white light collimated beam (b) Laser mono-chromatic light collimated beam. Fresnel diffraction and Poisson spot observed due to interference of light waves, more evident for the monochromatic light.}
\label{fig:figures5}
\end{figure*}

Detailed results and discussion are presented here on the effect of various system parameters on measurement accuracy.
The calibration target dots of known size and depth locations are captured in different background illumination configurations. The size estimated by the proposed technique is compared with the actual dot size at various depth locations. Measurements performed at a magnification of $\sim6.8$x are depicted in Figs.~\ref{fig:figures1}, \ref{fig:figures2}, and \ref{fig:figures3}. Two background intensity levels: low(0.2) and high(0.65) were considered. These are the rescaled average background image pixel bit value, where, for a 16-bit image, the pixel value ranges from 0-65535 and is rescaled to 0-1.
\\
\noindent \textbf{Diffused Background Illumination:}
 The size is accurately predicted within a relative error of 5-10\% within the measurement depth when using the diffused light source (Fig.~\ref{fig:figures2}a,c). The measurements of target dots further validate the linear relationship between $\sigma$ and $|\Delta z|$, as shown in Fig.~\ref{fig:figures3}a,c. The intensity of the background illumination is found to have no significant effect on the results. The use of diffused white light yields better results due to its incoherent nature.
\\
\noindent \textbf{Collimated Beam Illumination:}
 Measurements with a collimated light source exhibit a higher relative error, as shown in Fig.~\ref{fig:figures2}b,d. This is due to the interference pattern caused by Fresnel diffraction and Poisson spot formation as depicted in Fig.~\ref{fig:figures5}. The presence of interference patterns causes significant deviations in the gradients, which do not align with the expected profiles based on Gaussian PSFs. This non-Gaussian blurring of the dots invalidates the proposed analysis. However, in the case of a collimated white light beam, this error is prominent only for the smaller dots ($\leq 30\mu m$), since the interference patterns due to different wavelengths average out at length scales associated with the larger dot sizes. The measurements also deviate from the hypothesized linear relationship between $\sigma$ and $|\Delta z|$, as illustrated in Fig.~\ref{fig:figures3}b,d. However, the depth of detection is substantially increased, as evident from the higher values of $\alpha$. The background illumination intensity has minimal impact on the results.
\vspace{3mm}
\\
\noindent The second set of measurements, shown in Fig.~\ref{fig:figures4}, is performed using a diffused laser beam illumination at a higher magnification of approximately 13.7x. The same low and high normalised background intensity level as earlier is ensured for these measurements. For the particle of the same physical size, the higher magnification ensures the availability of more pixels to extract accurate information. Hence, with roughly twice as many pixels available in this second set,  a slightly better estimation of size is achieved than with the first set of measurements. The validity of the proposed technique is demonstrated for particle sizes as small as 7$\mu m$ with a suitable resolution.

\subsection*{Limits on Point Spread Function (PSF)}
The presumed Gaussian PSF in optical systems is limited by the diffraction of light waves and the formation of the Airy disk. This limits the resolution of the system as well as the validity of the proposed DFD approach. In Fig.~\ref{fig:figures5}, we have observed how interference patterns emerge owing to diffraction around the particle edges for the cases of collimated beams. In this section, we estimate the size of the smallest PSF i.e., Airy disk, to see if it affects the measurement analysis.
\\
For the given combination of lenses (Navitar $1.5\times$ lens attachment + $6.5\times$ Zoom lens + $1.0\times$ or $2.0\times$ adapter) being used for the parametric study using target dots, the Objective Numerical Aperture ${NA}_{obj}$ as given by the manufacturer is 
\begin{equation*}
    {NA}_{obj} = 0.106
\end{equation*}

\noindent The corresponding F-number $(f/\#)$ is given as
\begin{equation} \label{eq:fnum}
    (f/\#) = \frac{1}{2{NA}_{obj}}
\end{equation}

\noindent Then, Airy disk diameter $d_{Airy}$ in terms of $(f/\#)$ is given by \citep{stokseth_properties_1969}
\begin{equation} \label{eq:airy}
    d_{Airy} \approx 2.44 \lambda (f/\#)
\end{equation}

\noindent For the Cavilux light source $\lambda = 640nm$ (Red). Substituting values in Eq.~\ref{eq:fnum} and ~\ref{eq:airy} we get

\begin{equation*}
    d_{Airy} \approx 7.37 \mu m
\end{equation*}

\noindent Even in the case of a white light source, the components with a longer wavelength will form a larger Airy disk, as evident from Eq.~\ref{eq:airy}, and hence we can use the red light wavelength as a test case to evaluate the limitations.
\\
The least squared error fit of a Gaussian PSF to the Airy disk profile provides an equivalent Gaussian blurring standard deviation $\sigma_{eq}$ with respect to the airy disk diameter as  
\begin{equation} \label{eq:SD_eq}
    \sigma_{eq} \approx 0.127 \times d_{Airy}
\end{equation}

\noindent where the R-squared value of the fit is $R^2=0.9981$. Substituting the values in Eq.~\ref{eq:SD_eq}, we get
\begin{equation*}
    \sigma_{eq} = 0.9345 \mu m
\end{equation*}
\\
This value is smaller compared to the pixel size (refer to \textbf{Materials and Methods}, Section~\ref{sec:Materials and Methods} for details) and hence will not affect the results drastically. Furthermore, as a diffused light beam is suggested for the proposed technique, these diffraction effects will be significantly less obvious. 
\\
Although from Fig.~\ref{fig:figures3} and \ref{fig:figures4}c, one observes that the calculated blur kernel size approaches a finite non-zero value at focus $(|\Delta z|=0)$ instead of an expected sharp focused image with Dirac function as PSF (i.e., $\sigma=0$). This is expected due to the following reasons:

\begin{enumerate}
\item   The pixel intensity value is the average manifestation of the light intensity falling over the sensor. The image of a focused particle (both actual and artificial) has some pixels with intermediate intensity values at the boundary due to the edge of the projected shadow lying in an intermediate position within the pixel/sensor. This gives a sense of blurring even for the focused image with $\sigma \neq 0$.

\item   In theory, we need gradients at the edge of the particle image to approach infinity when in focus, which practically never seems to happen partially due to this discrete way of capturing information. The $\sigma$ calculations are further affected due to errors associated with the estimation of steep gradients from the available discrete information in the image.
\end{enumerate}

\section{Theoretical Particle Concentration Limit}\label{Appendix-A2}
\begin{figure*}[ht]
\centering
\includegraphics[width=0.8\linewidth]{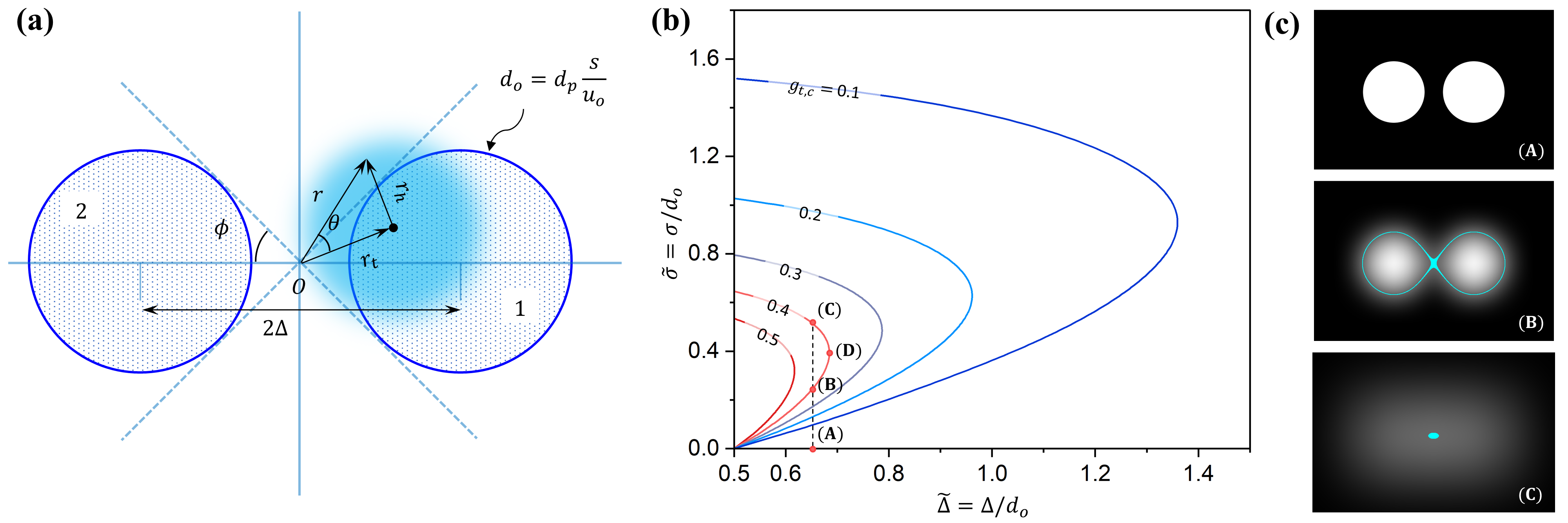}
\caption{\textbf{(a)} The blurred image for a particle pair is estimated by convolving the focused image with a Gaussian blur kernel, shown as a shaded circle. Here, 2$\Delta$ is the separation between the particles of the same size $d_p$. Intensity at point 'O' is estimated for different degrees of blur $\tilde{\sigma}$. $\phi$ is the angle that the tangent from 'O' makes with the horizontal axis. \textbf{(b)} Theoretical variation of non-dimensional parameter $\tilde{\sigma} = \sigma/d_o$  with inter-particle half separation $\tilde{\Delta} = \Delta/d_o$ for different intensity values ($g_{t,c}$) at the centre of the pair O. Two $\tilde{\sigma}$ solutions – near focus (B)  and far focus (C)  depth – exist for a prescribed $\tilde{\Delta}$ and $g_{t,c}$. Also, there is a critical separation $\tilde{\Delta}_c$ corresponding to (D) for a prescribed $g_{t,c}$  beyond which, for any depth, the particle pair is distinguishable. \textbf{(c)} Illustration of the blurred image of a particle pair corresponding to points (A), (B)  and (C) in (a). If $g_{t,c}$ exceeds the detection intensity threshold (0.4 here), then both particles cannot be directly distinguished, as shown. Here cyan represents $g_t=0.4 \pm 0.05$.}
\label{fig:figures6}
\end{figure*}

The proposed methodology is currently capable of analysing an isolated blurred particle. However, in sprays and other dispersed systems, particle images often overlap when projected along the optical axis onto the image plane. Blurring can cause particles to appear as a single indistinguishable non-symmetric entity, even if they do not overlap. The particle concentration limit is the extent to which the closely packed particles are distinguishable on the imaging plane based on a segmentation threshold value, which is chosen for the current study as $g_{t,c}=0.4$. This limiting condition is illustrated theoretically for a simple case where two particles of the same size $d_o$ are considered at a specified centre-to-centre distance of $2\Delta$, illustrated in Fig.~\ref{fig:figures6}a. The intensity at point 'O' is evaluated for different degrees of blurring $\widetilde{\sigma}$ and separation distance $\Delta$. If this exceeds the detection threshold value $g_{t,c}$, then the particles are indistinguishable. Convolution as earlier (Eq.~(\ref{eq:convolution})) is applied using a Gaussian blur kernel $h$ (Eq.~(\ref{eq:h_r})). In this case, the normalized image function $(i_f)$ takes a value of one within shaded regions (1) and (2) in Fig.~\ref{fig:figures6}a, and zero otherwise. These shaded regions can be defined geometrically in polar coordinates, with point 'O' as the origin

\begin{equation}
\begin{split}
    Region (1): & \quad r \geq \Delta \cos{\theta} - \sqrt{{r_o}^2-\Delta^2 \sin^2 {\theta}} \:\& \\
    & \quad r \leq \Delta \cos{\theta} + \sqrt{{r_o}^2-\Delta^2 \sin^2 {\theta}} \\
    & \quad -\phi \leq \theta \leq \phi
\end{split}
\end{equation}

\begin{equation}
\begin{split}
    Region (2): &  \quad r \geq -\Delta \cos{\theta} - \sqrt{{r_o}^2-\Delta^2 \sin^2 {\theta}} \:\& \\
    & \quad r \leq -\Delta \cos{\theta} + \sqrt{{r_o}^2-\Delta^2 \sin^2 {\theta}} \\
    & \quad \pi-\phi \leq \theta \leq \pi+\phi
\end{split}
\end{equation}
\\
Where $\phi$ in the angle subtended by the tangent to particle contour at origin as depicted in Fig.~\ref{fig:figures6}a. Substituting this into Eq.~(\ref{eq:convolution}) to evaluate $g_{t,c}$ at 'O' where $r_t=0$, while considering the additional non-dimensionalisation $\widetilde{\Delta} = \Delta/d_o$, we obtain the following expression:

\begin{multline}
    g_{t,c}=\frac{2}{\pi}\int_{-\phi}^{\phi}{\exp{\left(-\frac{1/4+{\widetilde{\Delta}}^2\cos{2\theta}}{2{\widetilde{\sigma}}^2}\right)}} \\
    \times \sinh{\left(\frac{2\widetilde{\Delta}\cos{\theta}\ \sqrt{1/4-{\widetilde{\Delta}}^2\sin^2{\theta}}}{2{\widetilde{\sigma}}^2}\right)}d\theta
\end{multline}

where $\phi=\sin^{-1}{\left(\frac{d_0}{2\Delta}\right)}=\sin^{-1}{\left(\frac{1}{2\widetilde{\Delta}}\right)}$. The solutions for this are numerically evaluated and variation of the dimensionless parameter $\widetilde{\sigma}$ with inter-particle half separation $\widetilde{\Delta}$ for different intensity values $\left(g_{t,c}\right)$ at the centre of the pair 'O' is depicted in Fig.~\ref{fig:figures6}b. Two solutions for $\widetilde{\sigma}$, at near focus depth and far focus depths, exist for a prescribed $\widetilde{\Delta}$ and $g_{t,c}$. Also, there is a critical separation ${\widetilde{\Delta}}_c$ for a prescribed $g_{t,c}$ beyond which, for any depth, the particle pair is distinguishable. For the chosen $g_{t,c}=0.4$ corrsponding to particle segmentation, this value is ${\widetilde{\Delta}}_c \approx 0.7$. This signifies the critical concentration limit, and particles with spacing such that $\widetilde{\Delta} > {\widetilde{\Delta}}_c$ are distinguishable for all depths in the measurement. In simpler terms, the particles with a spacing between their centres greater than 1.4 times the diameter will be distinguishable at all depths for the segmentation threshold of 0.4. 

This analysis is a simplified representation of the presence of such a limit to be considered when choosing the image based system for measurement. However, several further aspects must be considered when attempting to determine an absolute concentration limit for a given optical configuration. To start, most dispersed systems consist of multiple particles of different sizes, and the size distribution must be accounted for. Furthermore, there are two effects leading to overlap. Even if all particles were in the same plane perpendicular to the optical axis, the overlap would increase with the degree of out of focus, as treated above. This is very similar to the situation encountered in other out-of-focus approaches such as ILIDS/IPI, and concentration limits for such techniques have been derived previously \cite{damaschke_optical_2002}. However, with the DFD technique, we also encounter varying degrees of out of focus because the detection volume is also larger in the z-direction. This is an added influence that was not treated in the earlier work \cite{damaschke_optical_2002}. Finally, when attempting to determine a concentration limit theoretically, some assumption must be made regarding how uniform the concentration is throughout the detection volume, the most simple assumption being a uniform distribution.

\end{appendices}

\backmatter


\section*{Declarations}

\subsection*{Ethical Approval }
Not Applicable

\subsection*{Competing interests}
The authors report no competing interests.

\subsection*{Author's contributions}
SJR and CT conceptualized the methodology. SJR and SS performed the experiments and analysed the data. SJR developed the image processing code. SB and CT supervised the project. SJR and SS prepared the original draft, CT edited the draft, and all authors reviewed the manuscript.

\subsection*{Funding}
The financial support of the Science and Engineering Research Board of India is acknowledged in sponsoring author CT through the VAJRA Faculty scheme. SJR acknowledges the support from the Prime Minister’s Research Fellowship (PMRF).

\subsection*{Availability of data and materials}
The data that support the findings of this study are available from the corresponding author upon reasonable request.


\bibliography{sn-bibliography}

\end{document}